\documentstyle[preprint,aps]{revtex}

\tightenlines

\begin{document}

\draft

\title{\bf Series Expansions for three-dimensional QED}
\author{C.J. Hamer\cite{byline1}, J. Oitmaa\cite{byline2}, and
  Zheng Weihong\cite{byline3}} 
\address{School of Physics,    
The University of New South Wales,                                   
Sydney, NSW 2052, Australia.}                      

\date{Sept. 18, 1998}

\maketitle 

\begin{abstract}
Strong-coupling series expansions are calculated for the
Hamiltonian version of compact lattice electrodynamics
in (2+1) dimensions, with 4-component fermions.
Series are calculated for the ground-state energy per site, the 
chiral condensate, and the masses of `glueball' and
positronium states. Comparisons are made with
results obtained by other techniques.
\end{abstract}                    
\pacs{PACS Indices: 11.Ha.; 12.38Gc.\\ \\}


\narrowtext
\section{INTRODUCTION}

Quantum electrodynamics in 2+1 dimensions (QED$_3$) has
generated considerable interest over recent years.
The model is super-renormalizable, but shares a
number of important features with quantum chromodynamics
(QCD) in 3+1 dimensions: it is believed to be confining
at large distances (in the quenched approximation, at
least), while in the massless fermion limit
it displays a chiral-like symmetry which is
spontaneously broken\cite{cor80}.
It is thus an ideal laboratory for testing non-perturbative 
methods of analysis. Versions of the model may also
be relevant to theories of the new high-T$_c$ 
superconductors\cite{bas}. The version with two-component
massless fermions generates a dynamical mass for
the photon through a Chern-Simons term\cite{des}.
This complication can be avoided in the four-component
version\cite{cor80,bur87}, where ``chiral'' symmetry
is broken in the normal Goldstone fashion,
leading to a doublet of massless Goldstone bosons
analogous to the pion in QCD. For the four-component
model with $N_f$ flavours of massless fermions, 
there has been  a debate running for some time whether chiral
symmetry is broken for all values $N_f$\cite{cor80,pis,pen} or 
whether there is a critical value $N_c\simeq 3.5$ above which no
spontaneous symmetry-breaking takes place\cite{dag,app,gus}.
We shall have nothing to say about this question.

The four-component version has been studied by several
different techniques, but we remain far from a complete
understanding of the model. Euclidean lattice
Monte Carlo simulations have been performed by 
several groups\cite{dag,far,burkit}. A number of 
authors\cite{pis,pen,gus,bur92,bur91} have used Schwinger-Dyson
techniques to study the chiral symmetry breaking, and
Allen and Burden\cite{allen} have also produced
estimates of the bound-state meson spectrum at
finite fermion masses. The Hamiltonian lattice
version has been studied by means of strong-coupling
expansions\cite{bur} and a loop expansion 
technique\cite{aro}.
A light-front approach has also been discussed\cite{burk,tam};
and the non-relativistic limit has been analyzed in some
detail\cite{tam,cor,yun}.

Here we treat the Hamiltonian lattice model by using linked-cluster
techniques\cite{he90} to generate further strong-coupling
series, thus extending the previous results of 
Burden and Hamer\cite{bur} (hereafter referred
to as I). It is well-known that Euclidean Monte Carlo
techniques are difficult and expensive to apply to models
with dynamical fermions, and so it seems worthwhile
to see if other techniques such as strong-coupling
expansions can give useful information in such cases.
A previous analysis of the Schwinger model\cite{ham97}
did indeed show that strong-coupling series approximants
can converge well into the weak coupling region. While
not quite as accurate as the exact finite-lattice
technique, the series approach did give quantitative estimates
of the lowest bound-state mass in the continuum limit,
at about the 5-10\% level of accuracy.

The paper begins with an outline of the
lattice formulation of the model in section II,
followed by a brief summary of the methods of
calculation in Section III.
Our results are presented in Section IV, discussing the 
ground-state energy, the chiral condensate, the
``glueball'' masses, and the spectrum of the bound-state
mesons as a function of the bare fermion mass.
Our conclusions are summarized in Section V.

\section{Formalism}

\subsection{Continuum Formulation}

The continuum Lagrangian density takes the standard form
\begin{equation}
{\cal L} = - {1\over 4} F_{\mu \nu} F^{\mu \nu} + 
\bar{\psi} (i \not\!\partial - e \not\!\!A - m) \psi
\end{equation}
where
\begin{equation}
F_{\mu \nu} = \partial_{\mu} A_{\nu} - \partial_{\nu} A_{\mu}
\end{equation}
and the Lorentz indices $\mu, \nu=0$, 1 or 2. The electric coupling $e$ in
(2+1) dimensions has the dimensions of $({\rm mass})^{1/2}$. 
Choosing the timelike axial
gauge
\begin{equation}
A_0 =0
\end{equation}
the Hamiltonian is found to be
\begin{equation}
H = \int {\rm d}^2 x \{ - i \bar{\psi}
( \not\!\!\vec{\nabla} + i e \not\!\!\vec{A} ) \psi + m \bar{\psi} \psi
+ {1\over 2} (\vec{E}^2 +B^2 ) \} \label{eq4}
\end{equation}
where 
\begin{mathletters}
\begin{equation}
E^i = F^{i0}= - \dot{A}^i
\end{equation}
and
\begin{equation}
B = \partial_1 A^2 - \partial_2 A^1
\end{equation}
\end{mathletters}
Note that the magnetic field $B$ has only one component 
in (2+1)D. Here $\psi$ is taken as a single four-component
Dirac spinor\cite{pis}, and $\gamma_0$, $\gamma_1$, $\gamma_2$
are $4\times 4$ Dirac matrices, for which we shall use the Dirac
representation where necessary:
\begin{equation}
\gamma_0 = \left( \begin{array}{rr}
\sigma_3  &0 \\
0  &-\sigma_3 
\end{array}\right) , \quad
\gamma_1 = i \left( \begin{array}{cc}
\sigma_1  & 0\\
0 & -\sigma_1
\end{array}\right), \quad
\gamma_2 = i \left( \begin{array}{cc}
\sigma_2  & 0\\
0 & -\sigma_2
\end{array}\right)
\end{equation}

In the zero-mass limit, the Hamiltonian (\ref{eq4}) possesses
a global $U(2)$ ``chiral'' symmetry\cite{pis}, whose
Lie algebra is spanned by the matrices
\begin{equation}
I, \quad \gamma_4= \left( \begin{array}{rr}
0  &I \\
I  &0 
\end{array}\right) , \quad
\gamma_5 = i \left( \begin{array}{cc}
0 & -I \\
I & 0
\end{array}\right),
\end{equation}
and $\gamma_{45}=-i \gamma_4 \gamma_5$. This
symmetry is expected to be spontaneously broken\cite{pis},
which should be manifested by a non-zero value of the chiral condensate 
$\langle \bar{\psi}\psi \rangle_0$.

At large fermion masses, a non-relativistic analysis can be carried
out\cite{tam,cor,yun}. Cornwall and Tiktopoulos and
Sen\cite{cor} showed that if the divergences were regulated by giving
a mass $\nu$ to the photon, then at one-loop order the renormalized
self-mass of the fermion is
\begin{equation}
m_R = m + {e^2\over 4 \pi } \ln \left( {2m \over \nu } \right)
\end{equation}
while the potential due to one-photon exchange between the electron and positron is:
\begin{equation}
V(r) = - e^2 \int {d^2k\over (2\pi)^2} {e^{i \bar{k}\cdot\bar{r}}\over
{\vec k}^2+\nu^2} = {e^2\over 2 \pi} \left( \gamma + \ln {\nu r\over 2}\right)
+O(\nu^2r^2)
\end{equation}
where $\gamma$ is Euler's constant. Both quantities show logarithmic 
divergences, but these divergences cancel in the Schr\"{o}dinger equation for
the positronium bound states
\begin{equation}
E \Psi (\vec{r}) = [-{1\over m} \vec{\nabla}_r^2 + 2 ( m_R - m) + V(r)]
\psi(\vec{r})
= [ - {1\over m} \vec{\nabla}_r^2 + {e^2\over 2 \pi} (\gamma + \ln mr )
] \psi (\vec{r})
\end{equation}
Numerical solutions of this equation\cite{tam,cor,yun} give
the `binding energies' of the lowest positronium states as
\begin{mathletters}
\begin{eqnarray}
E_0^0 && = {e^2\over 2\pi} \left( 1.7968 - {1\over 2} \ln \left( 
{2g^2 \over m\pi} \right) \right) \label{eq11a} \\ 
E_1^0 &&= {e^2\over 2\pi} \left( 2.9323 - {1\over 2} \ln \left( 
{2g^2 \over m\pi}\right) \right) 
\end{eqnarray}
\end{mathletters}
for angular momentum $l=0$, and
\begin{equation}
E_0^1 = {e^2\over 2\pi} \left( 2.6566 - {1\over 2} \ln \left( 
{2g^2 \over m\pi}\right) \right) 
\end{equation}
for $l=1$. At leading order the binding energies are
independent of ``spin'', so that each of these energy 
levels should be four-fold degenerate in the four-component fermion
model.

\subsection{Lattice Formulation}

A `staggered' Hamiltonian lattice formulation of this model has been 
discussed in reference I.
The four components of the
continuum fermion field fit naturally onto the four sites of
a $2\times 2$ unit cell on the 2-dimensional spatial lattice, leading
to a lattice Hamiltonian as follows:
\begin{equation}
H= {g^2\over 2a} W \label{eq11}
\end{equation}
where
\begin{equation}
W = W_0 + y W_1 + y^2 W_2 \label{eq12}
\end{equation}
and\footnote{The fermion mass term given in reference I
had the wrong sign. As it turns out, that did not
affect the results for the quantities they calculated.}
\begin{mathletters}
\begin{eqnarray}
W_0 &&= W_e + W_{\mu}
= \sum_l E_l^2 + \mu \sum_{\vec{r}} (-1)^{r_1+r_2+1} \chi^{\dag}(\vec{r})
\chi (\vec{r})  \\
W_1 &&= \sum_{\vec{r},i} \eta_i (\vec{r}) [ \chi^{\dag}(\vec{r}) U_i (\vec{r})
\chi(\vec{r}+\hat{i}) + {\rm h.c.} ] \\
W_2 &&= - \sum_p (U_p + U_p^{\dag} )
\end{eqnarray}
\end{mathletters}
Here $\vec{r}=(r_1, r_2)$ labels the sites, $l$ the links, 
$p$ the plaquettes and $i=1,2$ the directions on a square two-dimensional
spatial lattice with spacing $a$. The dimensionless coupling $g$ and mass
parameter $\mu$ are defined in terms of their continuum counterparts
$e$ and $m$ by
\begin{equation}
g^2 = e^2 a \quad {\rm and} \quad \mu={2am\over g^2}={2m \over e^2}
\end{equation}
while $y=1/g^2$,
and $\eta_1(\vec{r}) = (-1)^{r_2+1}$, $\eta_2 (\vec{r}) =1$.
The term $W_e$ is the electric field term, $W_{\mu}$ is the
fermion mass term, $W_1$ is the fermion kinetic energy, and
$W_2$ is the magnetic field energy, involving the usual plaquette 
operator $U_p$.

The correspondences between the lattice fields and their continuum
counterparts are for gauge fields
\begin{mathletters}
\begin{eqnarray}
{e\over a} E_l &&~\to~ E^i (\vec{x}) \\
A_l &&~\to~ A^i (\vec{x})
\end{eqnarray}
\end{mathletters}
where the link operator
\begin{equation}
U_l = \exp [ iea A_l (\vec{r}) ]
\end{equation}
while for the fermion field components
\begin{equation}
{1\over 2\sqrt{2} a} 
 \left[ \begin{array}{cccc}
0  & -i & 0 & 1 \\
1  &  0 & -i & 0 \\
-i  &  0 & 1 & 0 \\
0  &  1 & 0 & -i 
\end{array}\right] 
\left[ \begin{array}{c}
\xi_1 \\
\xi_2 \\
\xi_3 \\
\xi_4
\end{array}\right] 
\to
\left[ \begin{array}{c}
\psi_1 \\
\psi_2 \\
\psi_3 \\
\psi_4
\end{array}\right] \label{eq19}
\end{equation}
where\cite{sus}
\begin{equation}
\xi(\vec{r}) = i^{r_1+r_2} \chi (\vec{r}) \label{eq20}
\end{equation}
and the components $1,\cdots,4$ are assigned to sites 
of the $2\times 2$ unit cell as shown in Figure 1.

The commutation relations between the lattice fields are
\begin{mathletters}
\begin{equation}
 [ E_l, U_{l^{\prime}} ] = U_l \delta_{ll^{\prime}} 
\end{equation}
\begin{equation}
 [ E_l, U_{l^{\prime}}^{\dag} ]
      = - U_l^{\dag} \delta_{ll^{\prime}} 
\end{equation}
\begin{equation}
 \{ \chi^{\dag} (\vec{r}), \chi (\vec{r}^{\prime}) \} = 
     \delta_{\vec{r},\vec{r}^{\prime}} 
\end{equation}
\begin{equation}
 [E_l, \chi (\vec{r})] = [E_l, \chi^{\dag} (\vec{r})]
 = [U_l, \chi (\vec{r})]= [U_l, \chi^{\dag} (\vec{r})] =0
\end{equation}
\end{mathletters}
With these correspondences, it can be shown\cite{bur}
that the lattice Hamiltonian (\ref{eq11}) reduces to the
continuum Hamiltonian (\ref{eq4}) in the naive continuum
limit $a\to 0$.

The introduction of the lattice breaks the U(2) ``chiral''
symmetry down to a discrete symmetry generated
by shifts of one lattice spacing\cite{bur}.
A unit shift in either the $x$ or $y$ direction
leaves the kinetic term in the Hamiltonian (\ref{eq11}) 
invariant, but alters the sign of the mass term. 
The corresponding continuum field transformations, 
from (19), are
\begin{mathletters}
\begin{equation}
\psi \to e^{i (\pi/2)\gamma_4} \psi
\end{equation}
or
\begin{equation}
\psi \to e^{i (\pi/2)\gamma_5} \psi
\end{equation}
\label{eq22}
\end{mathletters}
respectively.

\subsection{The Strong-Coupling Limit}
The Hamiltonian (\ref{eq11}) acts an a Fock space spanned by the 
usual strong-coupling basis\cite{kog}. With each link is associated
an integer electric flux $n_l$ such that 
$E_l \vert n_l \rangle = n_l \vert n_l \rangle$.
The operators $U_l$ and $U_l^{\dag}$ increase and
decrease the flux on link $l$ by one unit, respectively.
Each site of the lattice can be in one of two
fermionic states $\vert +\rangle$ or $\vert -\rangle$ 
obeying
\begin{mathletters}
\begin{eqnarray}
&\chi^{\dag} \vert - \rangle = \vert + \rangle,\quad
&\chi^{\dag} \vert + \rangle =0  \\
&\chi\vert - \rangle =0, \quad &\chi\vert + \rangle
= \vert - \rangle
\end{eqnarray}
\end{mathletters}
Consider first the massless theory, $\mu=0$.
In the strong-coupling limit, the 
variable $y=0$ and the Hamiltonian $W$ reduces to
$W_e$. The ground-state is then highly degenerate,
having flux $n_l=0$ on each link, but with the fermionic
state entirely arbitrary. This degeneracy is
broken at the next order by the kinetic term
$W_1$, leaving only two degenerate states $\vert A\rangle$
and $\vert B\rangle$ whose fermionic content is
\begin{equation}
\vert A\rangle = \cases { \vert + \rangle, &on odd sites;\cr
\vert - \rangle, &on even sites.\cr}
\end{equation}
and
\begin{equation}
\vert B\rangle = \cases { \vert - \rangle, &on odd sites;\cr
\vert + \rangle, &on even sites.\cr}
\end{equation}
The chiral shifts of equation (\ref{eq22}) map these two
states into each other. When the mass term $W_{\mu}$
is included, chiral symmetry is explicitly broken and
state $\vert B\rangle $ is favoured energetically.
We thus take $\vert B\rangle$ as the unperturbed 
strong-coupling ground state for both the massive
and massless cases, and interpret this as the state
with no fermion excitations present.

An excitation on an odd or even site creates a positively or
negatively charged fermion respectively, i.e.
a positron or electron.
The first-order perturbation $W_1$ creates or
destroys an electron-positron pair on neighbouring sites,
joined by a link of flux. The second-order perturbation term
$W_2$ creates or destroys a plaquette of flux.
Gauge invariance ensures that for any state obtained from
the unperturbed vacuum by application of the operators 
$W_1$ and $W_2$, the net flux from any site is equal
to the charge of the fermion at that site, i.e.,
Gauss' law is obeyed.

\subsection{Positronium States}
This theory is expected to display confinement\cite{cor}, and the
only fermionic states with finite energy are expected to be
electrically neutral ``positronium'' bound states.
In the strong-coupling limit, the lowest energy positronium states
consist of an electron-positron pair on neighbouring sites,
connected by a link of unit flux. There are eight translationally-invariant
states of this type, corresponding to the eight links in the
unit cell, and we need to identify the linear combinations of
these states which correspond to eigenstates of the lattice
symmetry operators. The corresponding procedure for meson
states in four-dimensional Euclidean lattice QCD has been
discussed by Golterman\cite{gol}.

The symmetry group of the lattice Hamiltonian (\ref{eq11}) is
composed of the following elements:

\subsubsection{Even translations}
\begin{equation}
\chi (\vec{r}) \to \chi (\vec{r}+2\hat{i}),\quad
U_j (\vec{r})\to U_j (\vec{r} + 2 \hat{i})
\end{equation}
This corresponds to spatial translational invariance
in the continuum model.

\subsubsection{Odd translations}
\begin{mathletters}
\begin{equation}
\chi (\vec{r}) \to  \chi (\vec{r}+\hat{1}),
\quad U_i(\vec{r}) \to U_i (\vec{r}+\hat{1})
\end{equation}
or
\begin{equation}
\chi (\vec{r}) \to (-1)^{r_1} \chi (\vec{r}+\hat{2}),
\quad  U_i(\vec{r}) \to U_i (\vec{r}+\hat{2})
\end{equation}
\end{mathletters}
These are the discrete lattice versions of ``chiral''
symmetry corresponding to equations (\ref{eq22}).
The massless Hamiltonian is symmetric under these
transformations, but not the massive one.

\subsubsection{Diagonal shift $D$}
A combination of a shift by one site in the $x$ direction and
one site in the $y$ direction gives a diagonal shift
\begin{equation}
\chi (\vec{r}) \to  (-1)^{r_1} \chi (\vec{r} + \hat{1}
+\hat{2}), \quad
U_i(\vec{r}) \to U_i (\vec{r} + \hat{1} + \hat{2})
\end{equation}
This corresponds to a discrete $\gamma_{45}$ rotation in the
continuum fields
\begin{equation}
\psi \to i \gamma_{45}\psi
\end{equation}
This remains a symmetry of the massive Hamiltonian also.

\subsubsection{Square lattice rotations, $R$}
Let $R$ denote a lattice rotation by $\pi/2$ about a perpendicular
axis, as shown by Fig. 2:
\begin{mathletters}
\begin{eqnarray}
\chi (\vec{r}) &&~\to~ R(\vec{r}\ ') \chi (\vec{r}\ ') \\
U_2 (\vec{r}) &&~\to~ U_1 (\vec{r}\ ') \\
U_1 (\vec{r}) &&~\to~ U_2^{\dag}(\vec{r}\ ' - \hat{2})
\end{eqnarray}
\end{mathletters}
where
\begin{equation}
r'_1=r_2, \quad r'_2=-r_1
\end{equation}
and
\begin{equation}
R(r_1, r_2 ) = {1\over 2} [ (-1)^{r_1} + 
(-1)^{r_2} + (-1)^{r_1 + r_2} -1 ]
\end{equation}
Repeated  rotations generate the rotational symmetry group of a square, 
with 4 elements. It corresponds to rotation in both space and
``spin'' in the continuum model.

\subsubsection{``Axial parity'' inversion, $A$}
The ``axial parity'' inversion is discussed by Burden and Allen\cite{bur92,allen}.
In the continuum, it corresponds to the operations:
\begin{equation}
\psi (x) ~\to ~ \psi '(x') = A \psi (x) , \quad
\bar{\psi}(x) ~\to ~ \bar{\psi}'(x') = \bar{\psi} (x) A^{-1}
\end{equation}
and the vector field transforms as:
\begin{equation}
\vec{A} (x) ~\to ~ \vec{A}'(x') = - \vec{A} (x) 
\end{equation}
where $x'=(x^0, -x^1, -x^2)$. A suitable representation
for the fermion operator $A$ is the matrix $i\gamma_0$.

On the lattice, this is simply
\begin{mathletters}
\begin{eqnarray}
&&\chi (\vec{r}) ~\to ~ \chi (-\vec{r}) \\
&&U_i  (\vec{r}) ~\to ~ U_i^{\dag} (-\vec{r} - \hat{i} ) \\
\end{eqnarray}
\end{mathletters}
which is equivalent to $R^2$, a rotation by $\Pi$ in
2+1 dimensions.

\subsubsection{Reflection, $\Pi$}
A reflection in the $y$ axis corresponds to 
\begin{mathletters}
\begin{eqnarray}
\chi (\vec{r} ) &&~\to~ \chi (\vec{r}\ ') \\
U_1 (\vec{r} ) &&~\to~ U_1^{\dag} (\vec{r}\ '-\hat{1}) \\
U_2 (\vec{r}) &&~\to~ U_2 (\vec{r}\ ')
\end{eqnarray}
\end{mathletters}
where 
\begin{equation}
r_1'=-r_1,\quad r_2'=-r_2
\end{equation}


\subsubsection{Charge conjugation, $C$}

The charge conjugation operation is also discussed by
Burden and Allen\cite{bur92,allen}. In the
continuum, it corresponds to the operators:
\begin{equation}
\psi(x) ~\to ~ \psi'(x) = C \bar{\psi}^T (x), \quad
\bar{\psi} (x) ~\to ~ \bar{\psi}' (x) = - \psi^T (x) C^{\dag}
\end{equation}
and
\begin{equation}
A^{\mu} (x) ~\to ~ A^{\mu\prime}(x) = - A^{\mu} (x)
\end{equation}
The fermion operator can be represented, up to an 
arbitrary phase, by the matrix $\gamma_2$. The 
translation to lattice variables is slightly involved,
and is best handled in terms of the fields $\xi_i$
defined in equations (\ref{eq19}), (\ref{eq20}).
We shall not go into further details here.


The translationally invariant positronium
eigenstates can be classified in terms of their
eigenvalues under these symmetry operations.
The group of square rotations is ${\cal C}_4$, the cyclic
group of 4 elements. It has 4 irreducible representations,
each with dimension 1. The allowed eigenvalues of $R$ simply
consist of the 4th. roots of unity (i.e. powers of
$\epsilon \equiv e^{- 2 \pi i/4}$).
The shift $D$, reflection $\Pi$ and axial
parity $A$ each generate 2-element groups, 
with eigenvalues $\pm 1$ for the positronium 
states with translational symmetry.


In the strong-coupling limit, the ``link excitations''
on the unit cell corresponding to the lowest energy
positronium states consist of the eight states listed in Table 1,
numbered according to Figure 3. These states transform into
each other under the action of the symmetry operators.
By taking linear 
combinations of these states, one can form eigenstates 
$\vert \psi_1\rangle, \cdots,\vert \psi_8 \rangle$
of the
lattice symmetry operators,  as listed in Table 2. Note
that the states with rotation eigenvalue $R=\epsilon = -i$
or $R=\epsilon^3=-i$ cannot simultaneously be eigenstates
of $\Pi$,  because a reflection converts $R=i$ to
$R=-i$, and vice versa. We have chosen to list states
 $\vert \psi_5\rangle$ to $\vert \psi_8\rangle$ which 
correspond to eigenstates of $\Pi$,
and are thus symmetric or antisymmetric combinations of the
states with $R=\pm i$.
Similarly, we have chosen to list states $\vert \psi_2 \rangle$
and $\vert \psi_4 \rangle$ which are eigenstates of $C$, rather
than $R$.
All eight states are degenerate in energy
in the strong-coupling limit $y\to 0$.
The fact that the Hamiltonian is symmetric under both rotations and
reflections implies that the pair $\vert \psi_5\rangle$ and
$\vert \psi_6\rangle$ will remain degenerate at all couplings,
and likewise the pair $\vert \psi_7 \rangle$ and $\vert \psi_8\rangle$.
The combination of rotation and charge conjugation symmetry
implies that the pair $\vert \psi_2\rangle $ and $\vert \psi_4\rangle$
will also remain degenerate.

In the `naive continuum limit' $a\to 0$, 
when $U_l \to 1 + iea A_l$, one finds that the quartet of
states $\vert \psi_5 \rangle$ to $\vert \psi_8 \rangle$
transcribe to simple combinations of quark and
antiquark fields on the lattice, and correspond to
`vector' states in the language of Burden and Allen\cite{bur92,allen},
with $J^{AC}= 1^{--}$, where $J$ is the `total angular
momentum'. The states $\vert \psi_1 \rangle$ to $\vert\psi_4\rangle$,
on the other hand, contain an admixture of gauge fields in the
naive continuum limit, and have no direct counterparts in the
catalogue of quark-antiquark states discussed by Burden and Allen.
The state  $\vert\psi_1\rangle$ is a scalar state, having the same
quantum numbers as the vacuum.


\subsection{Weak-coupling expansion}
Some useful information on the ground-state properties, at
least, can be gained by performing a ``weak-coupling'' expansion
for the lattice system as $y\to \infty$. For the present
model, this exercise was carried out in reference I.

The ground-state energy per site has an asymptotic expansion in the
weak-coupling limit:
\begin{equation}
\omega_0 \sim - 2 y^2 + 1.9162 y - {4 y\over \pi^2}
\int_0^{\pi/2} d q_1 \int_0^{\pi/2} d q_2 
[\cos^2 q_1 + \cos^2 q_2 + {\mu^2 \over 4 y^2} ]^{1/2}
+ O(1)\quad {\rm as} \quad y\to \infty  \label{eq37}
\end{equation}
where the integral arises from a sum over the
fermionic degrees of freedom. At very large $y$
one finds
\begin{equation}
\omega_0 \sim - 2 y^2 + 0.9581 y + O(1) \quad {\rm as} \quad
y\to \infty  \label{eq38}
\end{equation}
but for finite $y$ and $\mu$ it is more useful to
evaluate the integral as it stands.

The ground-state expectation value of the chiral condensate
was also calculated\cite{bur} as
\begin{equation}
\langle \bar{\psi}\psi \rangle^{\rm lattice}
\sim - {\mu\over \pi^2 y} \int_0^{\pi/2} dq_1
\int_0^{\pi/2} dq_2 [\cos^2 q_1 + \cos^2 q_2 + {\mu^2\over 4 y^2}
]^{-1/2} + O(y^{-2}) \quad {\rm as}\quad y\to\infty \label{eq39}
\end{equation}
where 
\begin{eqnarray}
\langle \bar{\psi}\psi \rangle^{\rm lattice}
&& = \langle \psi_0 \vert {1\over N} \sum_{\vec{r}}
(-1)^{r_1+r_2+1} \chi^{\dag} (\vec{r}) \chi (\vec{r}) 
\vert \psi_0\rangle \\
&& = {1\over e^4 y^2} \langle \bar{\psi}\psi \rangle^{\rm continuum}
\end{eqnarray}
in terms of the continuum chiral condensate. This quantity
is given by the Feynman Hellmann theorem as
\begin{equation}
\langle \bar{\psi} \psi\rangle^{\rm lattice}
= {1\over N} {\partial \omega_0\over \partial \mu}
\end{equation}
where $\omega_0$ is the ground-state eigenvalue of
$W$.

\subsection{Non-Relativistic Limit $m/e^2 \to \infty$}
When the fermion mass $m$ becomes very large, it should
be possible to study the model in a ``quenched'' approximation,
where fermion loop diagrams are suppressed. In a staggered
lattice formulation such as the present one, the suppression
of all fermion loops will lead to the ``static'' limit,
in which any fermion excitation is fixed at its initial
lattice site, and apart from the mass term the remaining lattice
Hamiltonian is simply that of the pure gauge theory.

To allow the fermions to move or migrate on the lattice, one
has to go to the next order in powers of $1/\mu$, the
mass parameter, and allow second-order diagrams involving
the excitation of an ($e^+e^-$) pair on neighbouring sites, as 
shown in Figure 4.
Figure 4a) represents a loop diagram in the vacuum sector,
and 4b) a similar diagram in the one-fermion sector; Figure
4c) illustrates a ``hopping'' diagram, in which an
existing fermion hops two lattice
sites. This allows fermion migration to take place.

It is useful to define new fermion variables representing the
excitations on the strong-coupling ground-state, namely:
\begin{equation}
\phi (\vec{r} )= \cases { \chi^{\dag} (\vec{r} ), &$(r_1+r_2)$ even;\cr
\chi (\vec{r} ), &$(r_1+r_2)$ odd.\cr}
\end{equation}
so that
\begin{equation}
\phi (\vec{r})\vert 0 \rangle = 0
\end{equation}
where $\vert 0 \rangle = \vert B\rangle $ is the strong-coupling
ground state. The Hamiltonian (\ref{eq12}) and link excitations 
(Table 1) are easily translated in terms of the new variables.

In the quenched approximation outlined above, the
effective Hamiltonian in the vacuum sector is
\begin{equation}
W^0_{\rm eff} = - {N\over 2}\mu - N {y^2\over \mu}
+ W_e + y^2 W_2
\end{equation}
(where $N$ is the  number of sites on the lattice) which is simply
the pure gauge field Hamiltonian plus a constant.
The first constant term is the negative energy of the
``Dirac sea'' on the lattice, while the second constant
term is the contribution of the second-order diagram,
Figure 4a). More interesting is the effective Hamiltonian
in the $(e^+e^-)$ sector, which takes the form:
\begin{equation}
W^{e^+e^-}_{\rm eff}  = - {(N-4)\over 2}\mu
- (N-4) {y^2\over \mu} + W_e + y^2 W_2  + y^2 W_1' 
\end{equation}
where
\begin{eqnarray}
W_1' = && {1 \over 2 \mu}  \sum_{\vec{r} {\rm ~odd}} 
(-1)^{\phi^\dag\phi (\vec{r})}
{\Big \{}
\phi^{\dag} (\vec{r}-\hat{1}) U_1^{\dag} (\vec{r}-\hat{1})
[ U_1^{\dag} (\vec{r}) \phi (\vec{r} +\hat{1}) + (-1)^{r_2+1}
( U_2 (\vec{r}-\hat{2})\phi(\vec{r}-\hat{2}) \nonumber \\
&& + U_2^{\dag} (\vec{r})
 \phi (\vec{r} + \hat{2} ))]
 + \phi^{\dag}(\vec{r}-\hat{2}) U_2^{\dag} (\vec{r}- \hat{2} ) [U_2 (\vec{r}) \phi (\vec{r}+ \hat{2} )
+ (-1)^{r_2+1} U_1^{\dag} (\vec{r}) \phi (\vec{r}+ \hat{1} )] \nonumber \\
&& + (-1)^{r_2+1} \phi^{\dag} (\vec{r}+ \hat{2}) U_2 (\vec{r}) U_1 (\vec{r}) \phi (\vec{r}+
\hat{1} ) + {\rm h.c.} {\Big \}} \nonumber \\
&& + {1\over 2 \mu} \sum_{\vec{r} {\rm ~even}}
(-1)^{\phi^{\dag} \phi (\vec{r})} {\Big \{}
\phi^{\dag} (\vec{r}- \hat{1} ) U_1 (\vec{r}-\hat{1} ) [
U_1 (\vec{r}) \phi (\vec{r}+ \hat{1} )\nonumber \\
&&  + (-1)^{r_2 + 1}
( U_2^{\dag} (\vec{r}- \hat{2} ) \phi (\vec{r}- \hat{2} ) + U_2 (\vec{r})
\phi (\vec{r}+ \hat{2} )] \nonumber \\
&& + \phi^{\dag}(\vec{r}-\hat{2} ) U_2 (\vec{r}-\hat{2} )
[ U_2^{\dag} (\vec{r}) \phi (\vec{r}+\hat{2} ) + (-1)^{r_2+1}
U_1 (\vec{r}) \phi (\vec{r}+\hat{1} ) ] \nonumber \\
&& + (-1)^{r_2+1} \phi^{\dag}(\vec{r}+\hat{2} ) U_2^{\dag} (\vec{r}) U_1^{\dag}
(\vec{r}) \phi (\vec{r}+ \hat{1} ) + {\rm h.c.} {\Big \} }
\end{eqnarray}

This clumsy expression is merely a constant, plus the pure gauge Hamiltonian,
plus ``hopping'' terms for the fermions in the six different paths allowed
for a double hop on the staggered lattice. The associated phase factor
\begin{equation}
(-1)^{\phi^{\dag}\phi(\vec{r})} \equiv  
\cases { +1, &if site $\vec{r}$ occupied;\cr
-1, &if site $\vec{r}$ unoccupied.\cr}
\end{equation}
accounts for the change of sign if the hopping fermion ``passes
through'' an occupied site.
These hopping terms correspond to the kinetic energy in the
non-relativistic continuum Hamiltonian.

The effective Hamiltonian is more complicated in form than the
original lattice Hamiltonian, but does not allow any
further fermion excitations, and would   therefore be quicker
and easier to implement in numerical calculations. We have not attempted
any such calculations as yet.

\section{Method}

To calculate the strong-coupling series for the model, we used Nickel's
cluster expansion method.
The techniques necessary were reviewed in
He et al \cite{he90}, and will not be repeated here.
In these calculations, the $W_0$  in Eq. (\ref{eq12}) 
 is taken as the unperturbed Hamiltonian, diagonal in the
basis of eigenvectors of $E_l$, while the  $W_1$ and $W_2$ in
Eq. (\ref{eq12})  then act as  perturbations.

To generate the  series for the ground state energy, we need to generate
a list of connected plaquette configurations, together with their lattice
constants and embedding constants. Since the first-order
perturbation $W_1$ and second-order perturbation $W_2$ 
involve links and plaquettes, respectively,
a cluster $\alpha$  will contribute terms $O(y^\alpha)$,
where $\alpha$ is given by
\begin{equation}
\alpha \ge 2 n_p + n_l \label{eqord}
\end{equation}
where $n_p$ is the number of plaquettes in $\alpha$, and $n_l$
the number of links not contained in plaquettes. Up to the order 
$y^{22}$ considered in the current paper, there is only one graph
(Figure 5) which does not obey the above  relation, it actually 
contributes to order $y^{22}$ (due to the combination of $W_2$ on each
plaquette and $W_1$ in outer links) instead
of $y^{24}$ according to Eq.(\ref{eqord}). There are a total of 5494
graphs which contribute up to order $y^{22}$ for the ground-state properties.

The calculation of  glueball masses involves a list of clusters, both
connected and disconnected, with at least one plaquette in 
each  graph. There are 457 graphs which contribute to order
$y^{10}$.

The calculation of meson masses generally involves  a list of both 
connected and disconnected clusters\cite{he90}.
The  eight different links in the unit cell (shown in Figure 3)
are not equivalent in those calculations, which means we cannot
identify clusters which are topologically equivalent, or
even use rotation or reflection symmetry. Thus the separate
clusters proliferate enormously in this case:
there are 164142 clusters contributing up to order $y^{10}$.
For the scalar meson mass $m_1$, the eight different bond types are equivalent, 
so we only need
actually  one bond type, and there are only 569 clusters which 
contribute up to order $y^{12}$.

\section{Results}

\subsection{Ground-state energy}

Using the linked-cluster expansion method, series have
been calculated for the ground-state energy per site
up to order $y^{22}$.
The first few terms are:
\begin{eqnarray}
\omega_0/N && = -{\mu\over 2} - {2 y^2\over 1+2 \mu} - {y^4\over 2}
+ {14 y^4\over (1+2\mu )^3} \nonumber \\
 && + {
y^6 (-4742 -5084 \mu -1640 \mu^2 -368 \mu^3 - 64 \mu^4
 )\over  (1+2 \mu)^5 (3 + 2 \mu) (7 + 2 \mu)} + 
 \cdots
\end{eqnarray}
where $\omega_0$ is the ground-state eigenvalue of
$W$.
The coefficients are listed for various fixed values of
the dimensionless mass parameter $\mu=2m/e^2$ in Table III.
These coefficients agree with those of reference I
up to $O(y^{16})$.

Extrapolating these series into the weak-coupling region using 
integrated differential approximants\cite{gut}, one obtains results as 
shown in Figure 6. For the large-mass case, $\mu=10$, it can be
seen that the strong-coupling approximants match on to the 
weak-coupling form (\ref{eq37}) very nicely at around $1/y\simeq 0.5$.
For the lower masses, the strong-coupling approximants do not converge
well enough to establish a precise matching, but
they are clearly quite consistent with the asymptotic
form (\ref{eq37}). It is noteworthy that the 
$\mu$-independent form (\ref{eq38})
is only attained for very weak couplings
$\mu^2/y^2 \lesssim 0.4$.

The successful matching between the strong-coupling
approximants  and the weak-coupling asymptotic form gives some confidence
 that the series coefficients
have been calculated correctly, and that the approximants converge
well enough to provide useful
information about the weak-coupling (continuum) behaviour.

\subsection{Chiral condensate}
A more interesting quantity is the chiral condensate.
The first few terms in the series are:
\begin{eqnarray}
\langle \bar{\psi} \psi \rangle^{\rm lattice} = && - {1\over 2} 
+ {4 y^2\over 1 + 2 \mu} - {84 y^4\over (1+2 \mu)^4} \nonumber \\
+ && { 8 y^6  (122987 + 245156\mu + 181668\mu^2 + 64656\mu^3 + 13392\mu^4 + 2048\mu^5 + 192\mu^6) \over 
(1 + 2\mu)^6(3 + 2\mu)^2(7 + 2\mu)^2 }+ \cdots
\end{eqnarray}
Further coefficients at fixed values of $\mu$ are listed in Table III.
Figure 7 shows the extrapolation of these series into the weak-coupling
region,  as compared with the weak-coupling
form (\ref{eq39}). Once again, the large mass results $(\mu=10)$ match
very well to the weak-coupling form, while the lower-mass ones have not yet attained
it before convergence of the strong-coupling approximants is lost.

The most interesting case is $\mu =0$, the zero mass limit. A graph of
$y^2 \langle \bar{\psi} \psi \rangle^{\rm lattice}$ against $1/y$ is shown in 
Figure 8. 
It can be seen that integrated differential approximants to the
series fail to converge below about $1/y \simeq 1.5$, although
Pad\'{e} approximants behave in a more consistent fashion.
If taken at face value, the Pad\'{e} approximants would
indicate a very large value of the chiral condensate in the
continuum limit $1/y\to 0$,
\begin{equation}
e^{-4} \langle \bar{\psi} \psi \rangle^{\rm physical} = -0.284(10)~,
\label{padechi}
\end{equation}
Also shown in Figure 8, however, are some Monte Carlo estimates
of the chiral condensate for the Euclidean version of this 
model\footnote{In making this comparison, we assumed
the Euclidean coupling $g_E$ and Hamiltonian coupling
$g_H$ are equal, lacking information on a more precise connection.}
by Burkitt and Irving\cite{burkit}.
Their results are roughly compatible with ours at about $1/y \simeq 0.6$,
but show a dramatic decrease in magnitude beyond that point.
Since this occurs well below the region of convergence of the
series approximants, the series provide little evidence
either to confirm or deny this phenomenon.


Another Monte Carlo calculation of the chiral condensate in
the non-compact version of the model has been carried out by
Dagotto, Kogut and Kocic\cite{dag}. They found a quite
different behaviour, in which 
$\beta^2 \langle \bar{\psi} \psi \rangle^{\rm lattice}$
(where $\beta=y=1/g^2$) plunges rapidly toward zero
at a low $\beta$ value around $\beta\simeq 0.4$, and
then levels out to a plateau at a very small value, 
around $\beta^2 \langle \bar{\psi} \psi \rangle \simeq 0.001$.
Although this value is only tracked to $\beta\simeq 1$,
they take this  as evidence of chiral symmetry breaking
in the continuum limit. Burden and Roberts\cite{bur91}
have also obtained a Schwinger-Dyson estimate of the
chiral order parameter,
\begin{equation}
e^{-4} \langle \bar{\psi} \psi \rangle \simeq -0.003
\end{equation}
which is at least a similar order of magnitude to
that of Dagotto {\it et al.}\cite{dag}. It 
differs by a full two orders of magnitude from the naive
extrapolation above, equation (\ref{padechi}).

It would be interesting to see a more extensive Monte
Carlo simulation of the compact lattice model, to check whether
the decrease seen by Burkitt and Irving\cite{burkit}
is real, and whether the chiral condensate 
subsequently levels out at a small plateau value as in the 
noncompact model. There seems to be very little prospect
that further series calculations could shed light
on these questions.

\subsection{`Glueball' masses}
Strong-coupling series for the `photonball' masses,
$m_A$ and $m_S$, corresponding in the strong-coupling limit to 
single plaquette excitations which  are antisymmetric
and symmetric under reflections, respectively, have been 
calculated to order $y^{10}$.
The leading terms for both series are:
\begin{equation}
m_{S,A} = 4 + 16 y^2/( (1+2 \mu) (1-2 \mu) (3+2 \mu) )+O(y^4)
\end{equation}
with the difference between the two series only emerging at 
order $y^4$. Further coefficients at fixed values of 
$\mu$ are listed in Table IV. The coefficients up to
$O(y^8)$ were previous calculated in reference I.

Approximants to these strong-coupling series are graphed in 
Figures 9 and 10, and compared with the results for the pure
gauge theory, which we have calculated previously\cite{ham92}  to
order $y^{32}$.
It can be seen that for large $\mu$ values the series behave very similarly
to the pure gauge case, and appear to be decreasing exponentially towards
zero as $y$ increases. 
Our results are consistent with those of Burkitt and Irving\cite{burkit},
who found a systematic downward shift in $m_A$ for moderate quark
masses, as seen for the case $\mu=2$ in Figs. 9 and 10. At
small $\mu$ values, however, the behaviour appears to
change somewhat. In the region $\mu\lesssim 1$, there is substantial
overlap and level crossing between the glueball and meson states
in the strong-coupling region, and the picture becomes
more confused. The series approximants show an initial
rise for the glueball masses in the strong-coupling
region, followed by an apparent turnover, at
least in the $\mu =0.5$ case, but the convergence is not 
sufficient to track the behaviour reliably into
the weak-coupling region.


\subsection{Positronium masses}
Strong-coupling series for the meson states discussed in Section
II have been calculated to $O(y^{10})$ (or to $O(y^{12})$
for $m_1$).
The first three terms for arbitrary $\mu$ are 
\begin{mathletters}
\begin{eqnarray}
m_1&=& 1 + 2 \mu + 14 y^2/(1+2 \mu) 
   - y^4 (535 + 186\mu + 60\mu^2 + 8\mu^3)/[3(1 + 2\mu)^3] +O(y^6)\\
m_2 &=& m_4= 1 + 2 \mu + 10 y^2/(1+2 \mu)
- y^4 (283 + 42\mu + 12\mu^2 + 8\mu^3)/[3(1 + 2\mu)^3] +O(y^6)\\
m_3 &=&  1 + 2 \mu + 6 y^2/(1+2 \mu)
+ y^4 (-175 + 6\mu + 36\mu^2 - 8\mu^3)/[3(1 + 2\mu)^3] +O(y^6)\\
m_5 &=& m_6 = 1 + 2 \mu + 6 y^2/(1+2 \mu)
- y^4 (199 + 66\mu + 12\mu^2 + 8\mu^3)/[3(1 + 2\mu)^3] +O(y^6)\\
m_7 &=& m_8 = 1 + 2 \mu + 6 y^2/(1+2 \mu)
- y^4 (199 + 66\mu + 12\mu^2 + 8\mu^3)/[3(1 + 2\mu)^3]+O(y^6)
\end{eqnarray}
\end{mathletters}
Further terms at  selected values of $\mu$ are given in Table V.

Figure 11 graphs the various meson masses as functions
of $1/y$, at a fixed, large mass parameter $\mu = 10$.
It can be seen that the series
approximants for these masses do not converge below
about $1/y\simeq 1.5$.
For the lowest vector meson states there is a suggestion
that the mass reaches a peak at $1/y\simeq 1.5$, 
and then turns downward. A crude linear extrapolation 
has been made to estimate the continuum limit, but
the uncertainly in the estimate is large.


The resulting values for the vector mass are graphed
as a function of $\mu$ in Figure 12, along with
the Schwinger-Dyson estimates of Allen and Burden\cite{allen},
and the non-relativistic prediction\cite{tam,cor,yun}.
It can be seen that the lattice estimates of the positronium
binding energy have large errors, 
especially at smaller $\mu$, 
and lie about three times higher than the Schwinger-Dyson
values. Neither the Schwinger-Dyson
nor the lattice estimates show any definite evidence of
the logarithmic increase at large $\mu$ predicted by
the non-relativistic theory.


Figure 13 shows the masses as functions of $1/y$
for the massless case $\mu=0$. Once again, 
convergence is lost at rather small $y$ values,
around $1/y\simeq 2.5$, and it is hardly possible
to make useful estimates of the continuum limit.
There is no sign of any of the masses dropping towards
zero, or acting like a Goldstone boson. This is
because the expected Goldstone boson states are
the ``axiscalar'' and ``axipseudoscalar'' states\cite{allen},
which are not among the single-link excitations in the
strong-coupling limit which we have treated here (see
Section IID). The Goldstone bosons probably correspond to
L-shaped double-link excitations, which will
transform into each other under a single plaquette excitation.
It would be interesting to study their behaviour, but
we have not yet attempted such a study, owing to
technical complications.

\section{Summary and Discussion}
New strong-coupling series have been presented for
the ground-state energy and chiral condensate,
along with the `glueball' and positronium masses, in 
4-component Hamiltonian lattice QED$_{\rm 2+1}$
with full dynamical fermions. This represents the
first attempt at a series calculation for the positronium
states in this model.

Two major features are evident from these results. Firstly,
there is a very clear separation of scales between the
`glueballs' and the positronium states as the continuum limit 
is approached. At large fermion mass $\mu$, the positronium
energies remain finite in the continuum limit,
while the `glueball' masses scale exponentially towards 
zero, presumably corresponding in the limit to massless
photon states. The same thing appears to happen at
smaller $\mu$ values, although our evidence for the exponential
decrease of the glueball mass is rather slim. This separation
of scales is consistent with earlier discussions of the
pure gauge model\cite{ref28}: in the
nonrelativistic or static fermion limit, if one sticks
to naive `engineering' dimension scales one will end up
with a theory of free, massless photons as above,
whereas if one renormalizes the scale as discussed
by Polyakov\cite{ref29} or G\"{o}pfert and
Mack\cite{ref30} one will obtain a theory of free,
{\it massive} bosons.

The second feature is that the addition of dynamical
fermions does not greatly affect the glueball masses at large $\mu$:
at $\mu=2$, for instance, the only effect was a reduction
of the glueball mass at fixed $y$ of order a percent or two.
For $\mu\lesssim 0.5$ the effect is more pronounced, however.

In general, the results of these calculations were somewhat
disappointing. The bulk ground-state energy per site
converged well enough into the weak-coupling region to
display a convincing match with analytic weak-coupling
expansions\cite{bur}, and the `glueball' masses converged
well enough to justify the statements given above.
Other quantities, however, were not mapped out with such success.
Series approximants to the chiral condensate
at $\mu=0$ only converged down to $1/y \simeq 1.5$, and were unable
to confirm or deny the rapid plunge in magnitude seen
by Burkitt and Irving\cite{burkit} around
$1/y\simeq 0.5$. It will require more detailed
Monte Carlo studies to confirm that this plunge
really occurs, and to determine whether the chiral
condensate then levels out
at a small but finite value.

Series approximants to the positronium masses likewise
only converged down to $1/y\simeq 1.5$, even at large mass
$\mu$, making any extrapolations to the continuum limit
very uncertain. Our estimates of the continuum `binding
energy' are thus of little more than qualitative
accuracy. They lie about three times higher than 
the Schwinger-Dyson estimates of
Allen and Burden\cite{allen}, which are not 
expected to be very accurate at large mass $\mu$
in any case. Neither
set of results shows any definite sign of the logarithmic
increase in energy with $\mu$ predicted by
the non-relativistic analyses\cite{tam,cor,yun}.
Once again, a detailed understanding of the 
positronium spectrum must await a more accurate
study, by Monte Carlo or other methods.

The question whether this model develops 
Goldstone bosons in the massless limit
$\mu=0$ was not explored. These `axiscalar'
and `axipseudoscalar' states\cite{allen}
are not among the single-link excitations
for which we have calculated strong-coupling
expansions. They are most likely to be found
among the `double-link' excitations.
It is technically more difficult to
calculate strong-coupling series for the
double-link excitations; but on the other hand,
the series may well converge  more quickly for these
low-lying excitations. This might provide
an interesting subject for further study.

In general, however, the prospects for further
series calculations look dim. In contrast to the 
Schwinger model case\cite{ham97}, the strong-coupling
series approximants do not converge far enough into the 
weak-coupling region to allow accurate extrapolations to
the continuum limit. Thus it seems there is still
no real alternative to Monte Carlo methods,
whatever their limitations, for the detailed 
investigation of lattice gauge theories in three and four dimensions.
The strong-coupling series provide an accurate `platform'
of results in the strong-coupling region,
which can provide useful calibration points for other methods;
but their extrapolation into the weak-coupling
region remains rather uncertain.

\acknowledgments
We are very grateful for helpful comments from Conrad Burden
and Alan Irving.
This work forms part of a research project supported by a grant 
from the Australian Research Council. 



\begin{figure}
\begin{picture}(1000,650)(-500,-150)
\put(-50,0){\line(3,0){400}}
\put(0,-50){\line(0,3){400}}
\put(-50,300){\line(3,0){400}}
\put(300,-50){\line(0,3){400}}
\put(50,50){1}
\put(330,50){2}
\put(50,330){4}
\put(330,330){3}
\multiput(0,0)(300,0){2}{\circle*{15}}
\multiput(0,300)(300,0){2}{\circle*{15}}
\put(0,0){\circle*{30}}
\put(-100,-100){$(r_1,r_2)=(0,0)$}
\end{picture}
\caption{Assignment of spinor components to sites of the
$2\times 2$ unit cell.
}
\label{fig:fig1}
\end{figure}
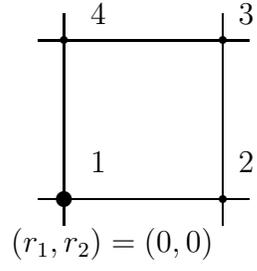

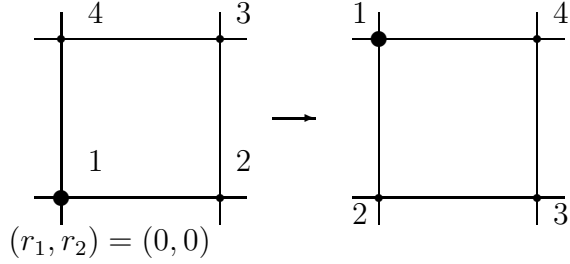
\begin{figure}
\begin{picture}(1200,700)(-200,-150)
\put(-50,0){\line(3,0){400}}
\put(0,-50){\line(0,3){400}}
\put(-50,300){\line(3,0){400}}
\put(300,-50){\line(0,3){400}}
\put(50,50){1}
\put(330,50){2}
\put(50,330){4}
\put(330,330){3}
\put(400,150){\vector(2,0){80}}
\put(550,0){\line(3,0){400}}
\put(600,-50){\line(0,3){400}}
\put(550,300){\line(3,0){400}}
\put(900,-50){\line(0,3){400}}
\put(550,-50){2}
\put(930,-50){3}
\put(550,330){1}
\put(930,330){4}
\multiput(0,0)(300,0){2}{\circle*{15}}
\multiput(0,300)(300,0){2}{\circle*{15}}
\multiput(0,0)(300,0){1}{\circle*{30}}
\multiput(600,0)(300,0){2}{\circle*{15}}
\multiput(600,300)(300,0){2}{\circle*{15}}
\multiput(600,300)(300,0){1}{\circle*{30}}
\put(-100,-100){$(r_1,r_2)=(0,0)$}
\end{picture}
\caption{A square lattice rotation by $\pi/2$. Point 1 is assumed
fixed.
}
\label{fig:fig2}
\end{figure}

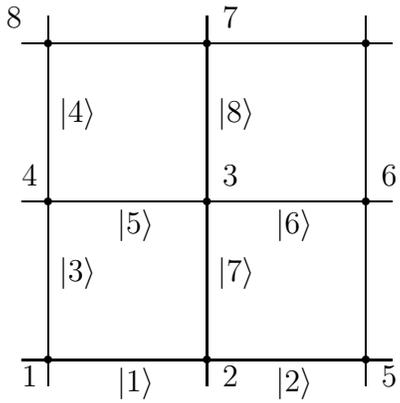
\begin{figure}
\begin{picture}(1200,1000)(-200,-100)
\put(-50,0){\line(3,0){700}}
\put(0,-50){\line(0,3){700}}
\put(-50,300){\line(3,0){700}}
\put(300,-50){\line(0,3){700}}
\put(-50,600){\line(3,0){700}}
\put(600,-50){\line(0,3){700}}
\put(-50,-50){1}
\put(330,-50){2}
\put(630,-50){5}
\put(-50,330){4}
\put(330,330){3}
\put(330,630){7}
\put(-80,630){8}
\put(630,330){6}
\put(130,-60){$\vert 1\rangle$}
\put(430,-60){$\vert 2\rangle$}
\put(130,240){$\vert 5\rangle$}
\put(430,240){$\vert 6\rangle$}
\put(20,150){$\vert 3\rangle$}
\put(20,450){$\vert 4\rangle$}
\put(320,150){$\vert 7\rangle$}
\put(320,450){$\vert 8\rangle$}
\multiput(0,0)(300,0){3}{\circle*{15}}
\multiput(0,300)(300,0){3}{\circle*{15}}
\multiput(0,600)(300,0){3}{\circle*{15}}
\end{picture}
\caption{Numbering of sites and ``link'' states on the unit
cell.
}
\label{fig:fig3}
\end{figure}

\begin{figure}
\begin{picture}(1000,600)(-200,-100)
\put(0,0){\framebox(200,400){}}
\put(0,0){\vector(0,2){200}}
\put(200,400){\vector(0,-2){200}}
\put(90,-80){a)}
\put(450,0){\line(0,3){400}}
\put(450,0){\vector(0,3){200}}
\put(500,0){\framebox(200,400){}}
\put(500,0){\vector(0,2){200}}
\put(700,400){\vector(0,-2){200}}
\put(580,-80){b)}
\put(1000,0){\line(0,3){300}}
\put(1000,0){\vector(0,3){100}}
\put(1000,300){\line(3,0){100}}
\put(1100,300){\line(0,-30){200}}
\put(1100,300){\vector(0,-30){100}}
\put(1100,100){\line(30,0){100}}
\put(1200,100){\line(0,30){300}}
\put(1200,100){\vector(0,30){150}}
\put(1100,-80){c)}
\end{picture}
\caption{Lattice perturbation theory diagrams at second
order involving excitation of an $(e^+e^-)$
pair. The associated gauge field excitations are
not shown. Horizontal lines represent the action of 
`link' excitation operators from $W_1$; vertical lines represent the
resulting fermion excitations.
}
\label{fig:fig4}
\end{figure}
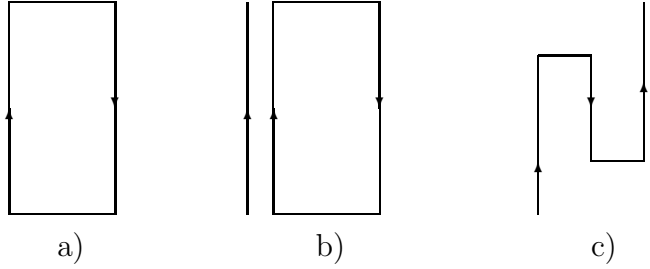

\begin{figure}
\begin{picture}(1000,600)(-200,-100)
\multiput(0,0)(200,0){3}{\framebox(200,200){}}
\multiput(0,200)(200,0){3}{\framebox(200,200){}}
\multiput(0,0)(200,0){4}{\circle*{15}}
\multiput(0,200)(200,0){4}{\circle*{15}}
\multiput(0,400)(200,0){4}{\circle*{15}}
\end{picture}
\caption{The last graphs contributing at order $y^{22}$.
}
\label{fig:fig5}
\end{figure}
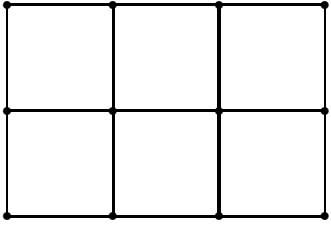

\begin{figure}
\caption{Graph of the ground-state energy per site,
 $y^{-2} \omega_0/N$ versus $1/y$,
for various fixed values of the mass parameter 
$\mu=0,0.5,2,10$. The curves at large $1/y$ are
integrated differential approximants to the strong-coupling series;
while those at small $1/y$ correspond to the asymptotic weak-coupling 
form (\ref{eq37}).
}
\label{fig:fig6}
\end{figure}

\begin{figure}
\caption{Graph of the chiral condensate,
 $ \langle \bar{\psi} \psi \rangle^{\rm lattice}$ versus $\mu/y$,
for various finite values of the mass parameter 
$\mu=0.5,2,10$. The curves at large $1/y$ are various integrated 
differential approximants and Pad\'{e} approximants to strong-coupling
series, while the solid line at small
$\mu/y$ is the weak-coupling asymptotic  form (\ref{eq39}).
}
\label{fig:fig7}
\end{figure}

\begin{figure}
\caption{Graph of the chiral condensate,
 $ y^2 \langle \bar{\psi} \psi \rangle^{\rm lattice}$ versus $1/y$,
for the massless case $\mu=0$.
The curves shown are integrated differential approximants
(solid lines) and 
$[n/(n+1)]$ Pad\'{e} approximants (broken lines) 
to the strong-coupling series. The last two 
Pad\'{e} approximants are
almost indistinguishable, and the successive intercepts at
$1/y=0$ are $0.232, 0.274, 0.283$ and 0.284.
}
\label{fig:fig8}
\end{figure}

\begin{figure}
\caption{The mass  $m_A$ as a function of $y$, for
$\mu =0,0.5,2,10$, Various  integrated 
differential approximants and Pad\'{e} approximants to strong-coupling series are
shown, together with corresponding results for the `pure gauge'
case ($\mu=\infty$).
}
\label{fig:fig9}
\end{figure}

\begin{figure}
\caption{The mass  $m_S$ as a function of $y$, as in Figure 9.
}
\label{fig:fig10}
\end{figure}

\begin{figure}
\caption{Strong-coupling series approximants to
$m/\mu$ as functions of $1/y$ for the meson states 
 $m_1,\cdots,m_8$, at mass parameter $\mu=10$.
}
\label{fig:fig11}
\end{figure}

\begin{figure}
\caption{Graph of $m/\mu$ versus $\mu$ for the vector state $m_7$.
}
\label{fig:fig12}
\end{figure}

\begin{figure}
\caption{Series approximants to
 $m_1,\cdots,m_8$ versus $1/y$ at  $\mu=0$.
}
\label{fig:fig13}
\end{figure}

\begin{table}
\caption{Link excitations on the unit cell 
corresponding to positronium states
in the strong-coupling limit, with sites labelled according to
Figure 3.
}\label{tab1}
\begin{tabular}{cccccc}
\multicolumn{1}{c}{Link state} &\multicolumn{1}{c}{Operator equivalent} \\
\hline
$\vert 1 \rangle $ & $\chi^{\dag} (2) U_1^{\dag} (1)\chi (1)$ \\
$\vert 2 \rangle $ & $\chi^{\dag} (2) U_1 (2)       \chi (5)$ \\
$\vert 3 \rangle $ & $\chi^{\dag} (4) U_2^{\dag} (1)\chi (1)$ \\
$\vert 4 \rangle $ & $\chi^{\dag} (4) U_2 (4)       \chi (8)$ \\
$\vert 5 \rangle $ & $\chi^{\dag} (4) U_1 (4)       \chi (3)$ \\
$\vert 6 \rangle $ & $\chi^{\dag} (6) U_1^{\dag} (3)\chi (3)$ \\
$\vert 7 \rangle $ & $\chi^{\dag} (2) U_2 (2)       \chi (3)$ \\
$\vert 8 \rangle $ & $\chi^{\dag} (7) U_2^{\dag} (3)\chi (3)$ \\
\end{tabular}
\end{table}

\begin{table}
\squeezetable
\caption{Linear combinations of the link states forming eigenstates of 
the lattice symmetry group ($\vert \psi_j\rangle
= a_i^{j} \vert i \rangle$), and their symmetry eigenvalues.
}\label{tab2}
\begin{tabular}{rrrrrrrrr}
\multicolumn{1}{c}{State:} &\multicolumn{1}{c}{$\vert \psi_1\rangle$} 
 &\multicolumn{1}{c}{$\vert \psi_2\rangle$} 
 &\multicolumn{1}{c}{$\vert \psi_3\rangle$} 
 &\multicolumn{1}{c}{$\vert \psi_4\rangle$} 
 &\multicolumn{1}{c}{$\vert \psi_5\rangle$} 
 &\multicolumn{1}{c}{$\vert \psi_6\rangle$} 
 &\multicolumn{1}{c}{$\vert \psi_7\rangle$} 
 &\multicolumn{1}{c}{$\vert \psi_8\rangle$} 
\\
\hline
\multicolumn{9}{c}{Amplitudes of the link states} \\
$a_1$ & 1 & 1 & 1 & 0 & 1 & 0 & 1 & 0 \\
$a_2$ & 1 & 1 & 1 & 0 &$-$1 & 0 &$-$1 & 0 \\
$a_3$ &$-$1 & 0 & 1 & 1 & 0 & 1 & 0 & 1 \\
$a_4$ &$-$1 & 0 & 1 & 1 & 0 &$-$1 & 0 &$-$1 \\
$a_5$ &$-$1 & 1 &$-$1 & 0 & 1 & 0 &$-$1 & 0 \\
$a_6$ &$-$1 & 1 &$-$1 & 0 &$-$1 & 0 & 1 & 0 \\
$a_7$ &$-$1 & 0 & 1 &$-$1 & 0 &$-$1 & 0 & 1 \\
$a_8$ &$-$1 & 0 & 1 &$-$1 & 0 & 1 & 0 &$-$1 \\
\hline
\multicolumn{9}{c}{Eigenvalues} \\
$R$ & +1 & $-$ & $-$1 & $-$ & $-$ & $-$ & $-$ & $-$ \\
$D$ & +1 & $-$1 & +1 & $-$1 & +1  & +1  & $-$1  & $-$1  \\
$\Pi$&+1 & +1 & +1 & +1 & $-$1  & +1  & $-$1  & +1  \\
$C$ & +1 & +1 & +1 & $-$1 & $-$1 & $-$1 & $-$1 & $-$1 \\
$A$ & +1 & +1 & +1 & +1 & $-$1 & $-$1 & $-$1 & $-$1 \\
\end{tabular}
\end{table}

\setdec 0.0000000000000
\begin{table}
\squeezetable
\caption{Series coefficients of $y^{2n}$ in strong-coupling
expansions of the ground-state energy $\omega_0$ and chiral condensate
$\langle \bar{\psi}\psi \rangle^{\rm lattice}$. 
}\label{tab3}
\begin{tabular}{rrrrrr}
\multicolumn{1}{c}{$n$} &\multicolumn{1}{c}{$\mu=0$}
&\multicolumn{1}{c}{$\mu=0.5$} &\multicolumn{1}{c}{$\mu=1$}
&\multicolumn{1}{c}{$\mu=2$} &\multicolumn{1}{c}{$\mu=10$} \\
\hline
\multicolumn{6}{c}{ground state energy}\\
  0 &\dec  0.000000000000 &\dec $-$2.500000000000$\times 10^{-1}$ &\dec $-$5.000000000000$\times 10^{-1}$ &\dec $-$1.000000000000 &\dec $-$5.000000000000 \\
  1 &\dec $-$2.000000000000 &\dec $-$1.000000000000 &\dec $-$6.666666666667$\times 10^{-1}$ &\dec $-$4.000000000000$\times 10^{-1}$ &\dec $-$9.523809523810$\times 10^{-2}$ \\
  2 &\dec  1.350000000000$\times 10^{1}$ &\dec  1.250000000000 &\dec  1.851851851852$\times 10^{-2}$ &\dec $-$3.880000000000$\times 10^{-1}$ &\dec $-$4.984882842026$\times 10^{-1}$ \\
  3 &\dec $-$2.258095238095$\times 10^{2}$ &\dec $-$7.562500000000 &\dec $-$1.088065843621 &\dec $-$1.057163636364$\times 10^{-1}$ &\dec $-$4.840190310415$\times 10^{-4}$ \\
  4 &\dec  4.740493349632$\times 10^{3}$ &\dec  4.168190104167$\times 10^{1}$ &\dec  2.828630278758 &\dec  1.440413481513$\times 10^{-1}$ &\dec  3.805417269032$\times 10^{-2}$ \\
  5 &\dec $-$1.145332120404$\times 10^{5}$ &\dec $-$2.641393774675$\times 10^{2}$ &\dec $-$8.251436202716 &\dec $-$1.196372180167$\times 10^{-1}$ &\dec  7.580050883728$\times 10^{-5}$ \\
  6 &\dec  3.019112271993$\times 10^{6}$ &\dec  1.828909211780$\times 10^{3}$ &\dec  2.673208303367$\times 10^{1}$ &\dec  1.528725434873$\times 10^{-1}$ &\dec $-$2.829817166242$\times 10^{-3}$ \\
  7 &\dec $-$8.446045864104$\times 10^{7}$ &\dec $-$1.343961734536$\times 10^{4}$ &\dec $-$9.195676938217$\times 10^{1}$ &\dec $-$2.146264820882$\times 10^{-1}$ &\dec $-$8.316158017118$\times 10^{-6}$ \\
  8 &\dec  2.467210478469$\times 10^{9}$ &\dec  1.031183072419$\times 10^{5}$ &\dec  3.302984237354$\times 10^{2}$ &\dec  3.075754366619$\times 10^{-1}$ &\dec $-$6.084851949294$\times 10^{-4}$ \\
  9 &\dec $-$7.447992091066$\times 10^{10}$ &\dec $-$8.175829204890$\times 10^{5}$ &\dec $-$1.226060361259$\times 10^{3}$ &\dec $-$4.580923967202$\times 10^{-1}$ &\dec $-$3.103254879714$\times 10^{-6}$ \\
 10 &\dec  2.307292550322$\times 10^{12}$ &\dec  6.651573204917$\times 10^{6}$ &\dec  4.670066518532$\times 10^{3}$ &\dec  6.989367972569$\times 10^{-1}$ &\dec  4.685274190981$\times 10^{-4}$ \\
 11 &\dec $-$7.298475917954$\times 10^{13}$ &\dec $-$5.525241939727$\times 10^{7}$ &\dec $-$1.816217581154$\times 10^{4}$ &\dec $-$1.087547736352 &\dec  2.784016199972$\times 10^{-6}$ \\
\hline
\multicolumn{6}{c}{chiral condensate}\\
  0 &\dec  5.000000000000$\times 10^{-1}$ &\dec  5.000000000000$\times 10^{-1}$ &\dec  5.000000000000$\times 10^{-1}$ &\dec  5.000000000000$\times 10^{-1}$ &\dec  5.000000000000$\times 10^{-1}$ \\
  1 &\dec $-$4.000000000000 &\dec $-$1.000000000000 &\dec $-$4.444444444444$\times 10^{-1}$ &\dec $-$1.600000000000$\times 10^{-1}$ &\dec $-$9.070294784581$\times 10^{-3}$ \\
  2 &\dec  8.400000000000$\times 10^{1}$ &\dec  5.250000000000 &\dec  1.037037037037 &\dec  1.344000000000$\times 10^{-1}$ &\dec  4.319187992657$\times 10^{-4}$ \\
  3 &\dec $-$2.231056689342$\times 10^{3}$ &\dec $-$3.661718750000$\times 10^{1}$ &\dec $-$3.414650205761 &\dec $-$1.856045336482$\times 10^{-1}$ &\dec $-$1.490233956148$\times 10^{-4}$ \\
  4 &\dec  6.534158638592$\times 10^{4}$ &\dec  2.809490017361$\times 10^{2}$ &\dec  1.222412306072$\times 10^{1}$ &\dec  2.632452778514$\times 10^{-1}$ &\dec  1.645776229220$\times 10^{-5}$ \\
  5 &\dec $-$2.025737847176$\times 10^{6}$ &\dec $-$2.284382049423$\times 10^{3}$ &\dec $-$4.639999014178$\times 10^{1}$ &\dec $-$3.921048572445$\times 10^{-1}$ &\dec  2.005503745248$\times 10^{-5}$ \\
  6 &\dec  6.518294767679$\times 10^{7}$ &\dec  1.929013101510$\times 10^{4}$ &\dec  1.831818873072$\times 10^{2}$ &\dec  6.165298249176$\times 10^{-1}$ &\dec $-$2.127283455260$\times 10^{-6}$ \\
  7 &\dec $-$2.153203741629$\times 10^{9}$ &\dec $-$1.672676257346$\times 10^{5}$ &\dec $-$7.429933326106$\times 10^{2}$ &\dec $-$9.978195968035$\times 10^{-1}$ &\dec $-$2.070374769933$\times 10^{-6}$ \\
  8 &\dec  7.252977297783$\times 10^{10}$ &\dec  1.479158204994$\times 10^{6}$ &\dec  3.074244834593$\times 10^{3}$ &\dec  1.648612396832 &\dec  1.799802187362$\times 10^{-7}$ \\
  9 &\dec $-$2.480301602121$\times 10^{12}$ &\dec $-$1.327976394916$\times 10^{7}$ &\dec $-$1.291627245890$\times 10^{4}$ &\dec $-$2.768670289809 &\dec $-$9.300630428741$\times 10^{-7}$ \\
 10 &\dec  8.584509655365$\times 10^{13}$ &\dec  1.206675388588$\times 10^{8}$ &\dec  5.492904223987$\times 10^{4}$ &\dec  4.708649820309 &\dec  1.110110045268$\times 10^{-7}$ \\
 11 &\dec $-$3.000444314331$\times 10^{15}$ &\dec $-$1.107246112597$\times 10^{9}$ &\dec $-$2.359089145008$\times 10^{5}$ &\dec $-$8.090428010991 &\dec  7.857720150141$\times 10^{-7}$ \\
\end{tabular}
\end{table}

\setdec 0.0000000000000
\begin{table}
\squeezetable
\caption{Series coefficients of $y^{2n}$ in strong-coupling
expansions of the symmetric and antisymmetric glueball mass gaps 
$m_S$ and $m_A$ of the dimensionless Hamiltonian $W$.
}\label{tab4}
\begin{tabular}{rrrrrr}
\multicolumn{1}{c}{$n$} &\multicolumn{1}{c}{$\mu=0$}
&\multicolumn{1}{c}{$\mu=0.5$} &\multicolumn{1}{c}{$\mu=1$}
&\multicolumn{1}{c}{$\mu=2$} &\multicolumn{1}{c}{$\mu=10$} \\
\hline
\multicolumn{6}{c}{symmetric glueball mass gaps $m_S$}\\
  0 &\dec  4.000000000000 &\dec  4.000000000000 &\dec  4.000000000000 &\dec  4.000000000000 &\dec  4.000000000000 \\
  1 &\dec  5.333333333333 &\dec  1.500000000000 &\dec $-$1.066666666667 &\dec $-$1.523809523810$\times 10^{-1}$ &\dec $-$1.743489157677$\times 10^{-3}$ \\
  2 &\dec $-$4.189814814815$\times 10^{1}$ &\dec $-$3.135416666667 &\dec  4.769703703704 &\dec $-$2.352469495735$\times 10^{-1}$ &\dec $-$4.165631870342$\times 10^{-1}$ \\
  3 &\dec  5.935307344636$\times 10^{2}$ &\dec $-$1.170910493827$\times 10^{-2}$ &\dec $-$2.394715197442$\times 10^{1}$ &\dec $-$4.890552931893$\times 10^{-1}$ &\dec $-$3.937708897154$\times 10^{-4}$ \\
  4 &\dec $-$3.229319000985$\times 10^{3}$ &\dec  8.663828760713$\times 10^{1}$ &\dec  1.501264118830$\times 10^{2}$ &\dec  2.073466005647 &\dec  4.945486284578$\times 10^{-1}$ \\
  5 &\dec $-$6.430232339023$\times 10^{5}$ &\dec $-$1.617503042067$\times 10^{3}$ &\dec $-$8.784848865979$\times 10^{2}$ &\dec $-$5.575125327898 &\dec  1.071338323552$\times 10^{-3}$ \\
\hline
\multicolumn{6}{c}{antisymmetric glueball mass gaps $m_A$}\\
  0 &\dec  4.000000000000 &\dec  4.000000000000 &\dec  4.000000000000 &\dec  4.000000000000 &\dec  4.000000000000 \\
  1 &\dec  5.333333333333 &\dec  1.500000000000 &\dec $-$1.066666666667 &\dec $-$1.523809523810$\times 10^{-1}$ &\dec $-$1.743489157677$\times 10^{-3}$ \\
  2 &\dec $-$4.306481481481$\times 10^{1}$ &\dec $-$4.302083333333 &\dec  3.603037037037 &\dec $-$1.401913616240 &\dec $-$1.583229853701 \\
  3 &\dec  6.055105324434$\times 10^{2}$ &\dec  1.838348765432$\times 10^{-2}$ &\dec $-$2.776504371231$\times 10^{1}$ &\dec $-$3.473208512644$\times 10^{-1}$ &\dec $-$1.358230163160$\times 10^{-3}$ \\
  4 &\dec  9.053923337359$\times 10^{3}$ &\dec  5.354226523059$\times 10^{1}$ &\dec  1.574287286646$\times 10^{2}$ &\dec  1.695152914866 &\dec  1.162208030611 \\
  5 &\dec $-$1.377596830177$\times 10^{6}$ &\dec $-$5.970881550866$\times 10^{2}$ &\dec $-$9.103588055806$\times 10^{2}$ &\dec $-$7.260136671150$\times 10^{-1}$ &\dec  2.449261377761$\times 10^{-3}$ \\
\end{tabular}
\end{table}

\setdec 0.0000000000000
\begin{table}
\squeezetable
\caption{Series coefficients of $y^{2n}$ in strong-coupling
expansions of the meson mass gaps 
$m_i$ $(i=1,2,\cdots,8)$ of the dimensionless Hamiltonian $W$.
}\label{tab5}
\begin{tabular}{rrrrrr}
\multicolumn{1}{c}{$n$} &\multicolumn{1}{c}{$m_1$}
&\multicolumn{1}{c}{$m_2=m_4$} &\multicolumn{1}{c}{$m_3$}
&\multicolumn{1}{c}{$m_5=m_6$} &\multicolumn{1}{c}{$m_7=m_8$} \\
\hline
\multicolumn{6}{c}{$\mu=0$}\\
  0 &\dec  1.000000000000 &\dec  1.000000000000 &\dec  1.000000000000 &\dec  1.000000000000 &\dec  1.000000000000 \\
  1 &\dec  1.400000000000$\times 10^{1}$ &\dec  1.000000000000$\times 10^{1}$ &\dec  6.000000000000 &\dec  6.000000000000 &\dec  6.000000000000 \\
  2 &\dec $-$1.783333333333$\times 10^{2}$ &\dec $-$9.433333333333$\times 10^{1}$ &\dec $-$5.833333333333$\times 10^{1}$ &\dec $-$6.633333333333$\times 10^{1}$ &\dec $-$6.633333333333$\times 10^{1}$ \\
  3 &\dec  3.625151675485$\times 10^{3}$ &\dec  1.757293650794$\times 10^{3}$ &\dec  1.200151675485$\times 10^{3}$ &\dec  1.341849206349$\times 10^{3}$ &\dec  1.253207231041$\times 10^{3}$ \\
  4 &\dec $-$9.101099254509$\times 10^{4}$ &\dec $-$4.164808339711$\times 10^{4}$ &\dec $-$2.976672075609$\times 10^{4}$ &\dec $-$3.324256060799$\times 10^{4}$ &\dec $-$3.008509651089$\times 10^{4}$ \\
  5 &\dec  2.534993350264$\times 10^{6}$ &\dec  1.113036398112$\times 10^{6}$ &\dec  8.177458288471$\times 10^{5}$ &\dec  9.143840136437$\times 10^{5}$ &\dec  8.115398963424$\times 10^{5}$ \\
  6 &\dec $-$7.521807853058$\times 10^{7}$ & & & & \\
\hline
\multicolumn{6}{c}{$\mu=0.5$}\\
  0 &\dec  2.000000000000 &\dec  2.000000000000 &\dec  2.000000000000 &\dec  2.000000000000 &\dec  2.000000000000 \\
  1 &\dec  7.000000000000 &\dec  5.000000000000 &\dec  3.000000000000 &\dec  3.000000000000 &\dec  3.000000000000 \\
  2 &\dec $-$2.683333333333$\times 10^{1}$ &\dec $-$1.283333333333$\times 10^{1}$ &\dec $-$6.833333333333 &\dec $-$9.833333333333 &\dec $-$9.833333333333 \\
  3 &\dec  1.365322916667$\times 10^{2}$ &\dec  6.154340277778$\times 10^{1}$ &\dec  4.075451388889$\times 10^{1}$ &\dec  5.498784722222$\times 10^{1}$ &\dec  4.683784722222$\times 10^{1}$ \\
  4 &\dec $-$8.920838363922$\times 10^{2}$ &\dec $-$3.876159660218$\times 10^{2}$ &\dec $-$2.760344845403$\times 10^{2}$ &\dec $-$3.678279451885$\times 10^{2}$ &\dec $-$2.870075748181$\times 10^{2}$ \\
  5 &\dec  6.436060668280$\times 10^{3}$ &\dec  2.730071323411$\times 10^{3}$ &\dec  2.011538052494$\times 10^{3}$ &\dec  2.664691694041$\times 10^{3}$ &\dec  2.110416878518$\times 10^{3}$ \\
  6 &\dec $-$4.984333949987$\times 10^{4}$ & & & & \\
\hline
\multicolumn{6}{c}{$\mu=1$}\\
  0 &\dec  3.000000000000 &\dec  3.000000000000 &\dec  3.000000000000 &\dec  3.000000000000 &\dec  3.000000000000 \\
  1 &\dec  4.666666666667 &\dec  3.333333333333 &\dec  2.000000000000 &\dec  2.000000000000 &\dec  2.000000000000 \\
  2 &\dec $-$9.740740740741 &\dec $-$4.259259259259 &\dec $-$1.740740740741 &\dec $-$3.518518518519 &\dec $-$3.518518518519 \\
  3 &\dec  2.117987280210$\times 10^{1}$ &\dec  9.052263374486 &\dec  5.541189674523 &\dec  9.760905349794 &\dec  6.595922184811 \\
  4 &\dec $-$7.103051896801$\times 10^{1}$ &\dec $-$2.746737478771$\times 10^{1}$ &\dec $-$1.941553551960$\times 10^{1}$ &\dec $-$3.281454899895$\times 10^{1}$ &\dec $-$2.157173735275$\times 10^{1}$ \\
  5 &\dec  2.054506982964$\times 10^{2}$ &\dec  9.091443816105$\times 10^{1}$ &\dec  6.590006451270$\times 10^{1}$ &\dec  1.138794166995$\times 10^{2}$ &\dec  7.209915733050$\times 10^{1}$ \\
  6 &\dec $-$8.384073194275$\times 10^{2}$ & & & & \\
\hline
\multicolumn{6}{c}{$\mu=2$}\\
  0 &\dec  5.000000000000 &\dec  5.000000000000 &\dec  5.000000000000 &\dec  5.000000000000 &\dec  5.000000000000 \\
  1 &\dec  2.800000000000 &\dec  2.000000000000 &\dec  1.200000000000 &\dec  1.200000000000 &\dec  1.200000000000 \\
  2 &\dec $-$3.229333333333 &\dec $-$1.277333333333 &\dec $-$2.213333333333$\times 10^{-1}$ &\dec $-$1.181333333333 &\dec $-$1.181333333333 \\
  3 &\dec  3.119580472860 &\dec  8.464108513709$\times 10^{-1}$ &\dec  4.398750760351$\times 10^{-1}$ &\dec  1.428440057720 &\dec  9.870369807970$\times 10^{-1}$ \\
  4 &\dec $-$4.249661491769 &\dec $-$1.098495894429 &\dec $-$8.941846899782$\times 10^{-1}$ &\dec $-$2.414224882476 &\dec $-$1.183255224582 \\
  5 &\dec  9.281773558871 &\dec  1.580697986782 &\dec  1.236101545345 &\dec  4.024451851449 &\dec  2.130785627667 \\
  6 &\dec $-$2.667077094545$\times 10^{1}$ & & & & \\
\hline
\multicolumn{6}{c}{$\mu=10$}\\
  0 &\dec  2.100000000000$\times 10^{1}$ &\dec  2.100000000000$\times 10^{1}$ &\dec  2.100000000000$\times 10^{1}$ &\dec  2.100000000000$\times 10^{1}$ &\dec  2.100000000000$\times 10^{1}$ \\
  1 &\dec  6.666666666667$\times 10^{-1}$ &\dec  4.761904761905$\times 10^{-1}$ &\dec  2.857142857143$\times 10^{-1}$ &\dec  2.857142857143$\times 10^{-1}$ &\dec  2.857142857143$\times 10^{-1}$ \\
  2 &\dec $-$5.901090594968$\times 10^{-1}$ &\dec $-$3.564409890941$\times 10^{-1}$ &\dec $-$1.625094482237$\times 10^{-1}$ &\dec $-$3.620559334845$\times 10^{-1}$ &\dec $-$3.620559334845$\times 10^{-1}$ \\
  3 &\dec $-$2.582504019689$\times 10^{-3}$ &\dec $-$1.053385459541$\times 10^{-2}$ &\dec  1.090172643379$\times 10^{-2}$ &\dec  3.819111465784$\times 10^{-2}$ &\dec  1.199822627856$\times 10^{-2}$ \\
  4 &\dec  2.457235107452$\times 10^{-1}$ &\dec  1.646791697513$\times 10^{-1}$ &\dec  7.346442898971$\times 10^{-2}$ &\dec  1.373283015502$\times 10^{-1}$ &\dec  1.747678732566$\times 10^{-1}$ \\
  5 &\dec  6.900777782581$\times 10^{-2}$ &\dec  1.213278624644$\times 10^{-2}$ &\dec $-$7.752772094654$\times 10^{-3}$ &\dec  5.188958501563$\times 10^{-3}$ &\dec $-$8.047408934744$\times 10^{-3}$ \\
  6 &\dec $-$1.752056769026$\times 10^{-1}$ & & & & \\
\end{tabular}
\end{table}

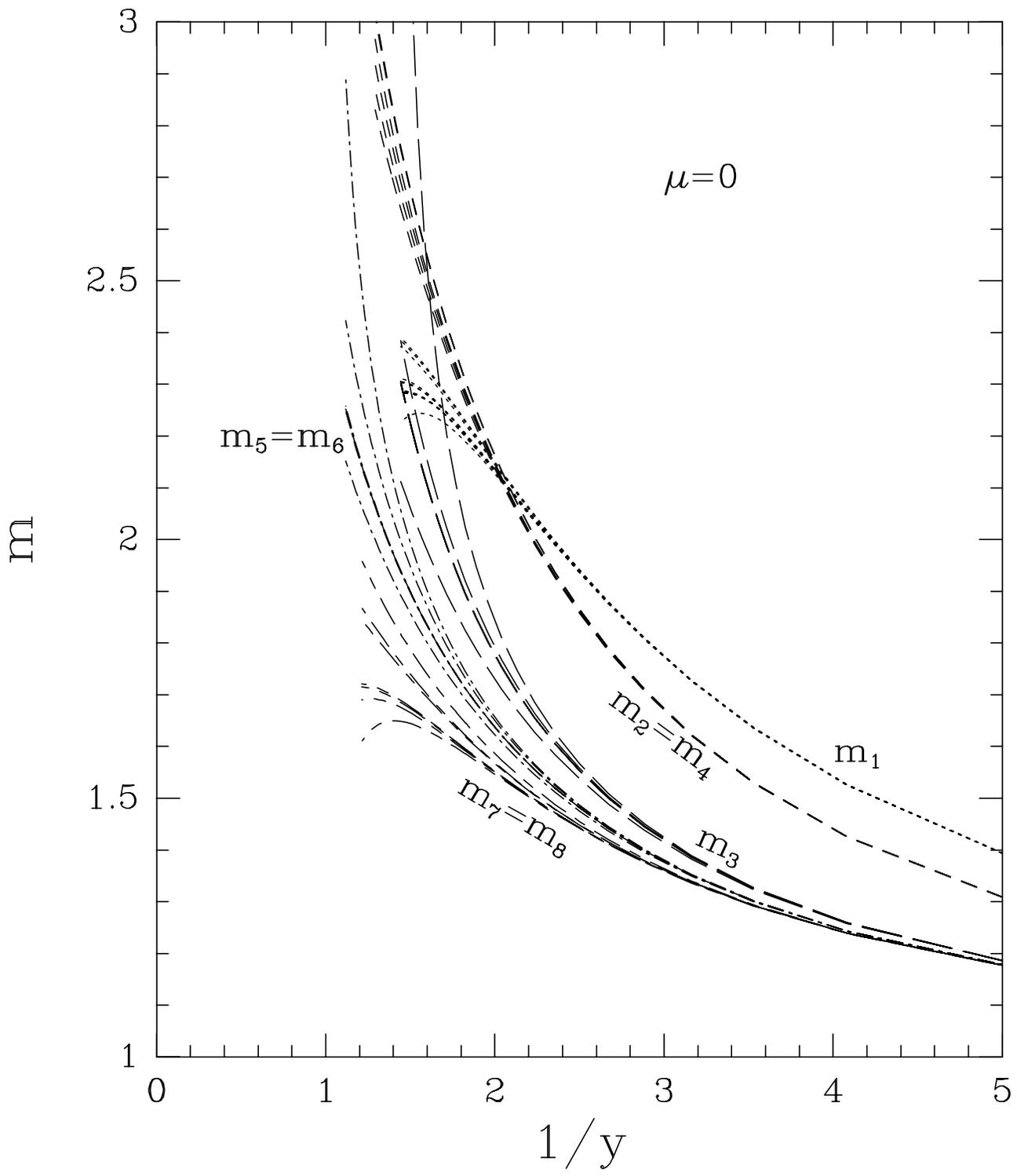
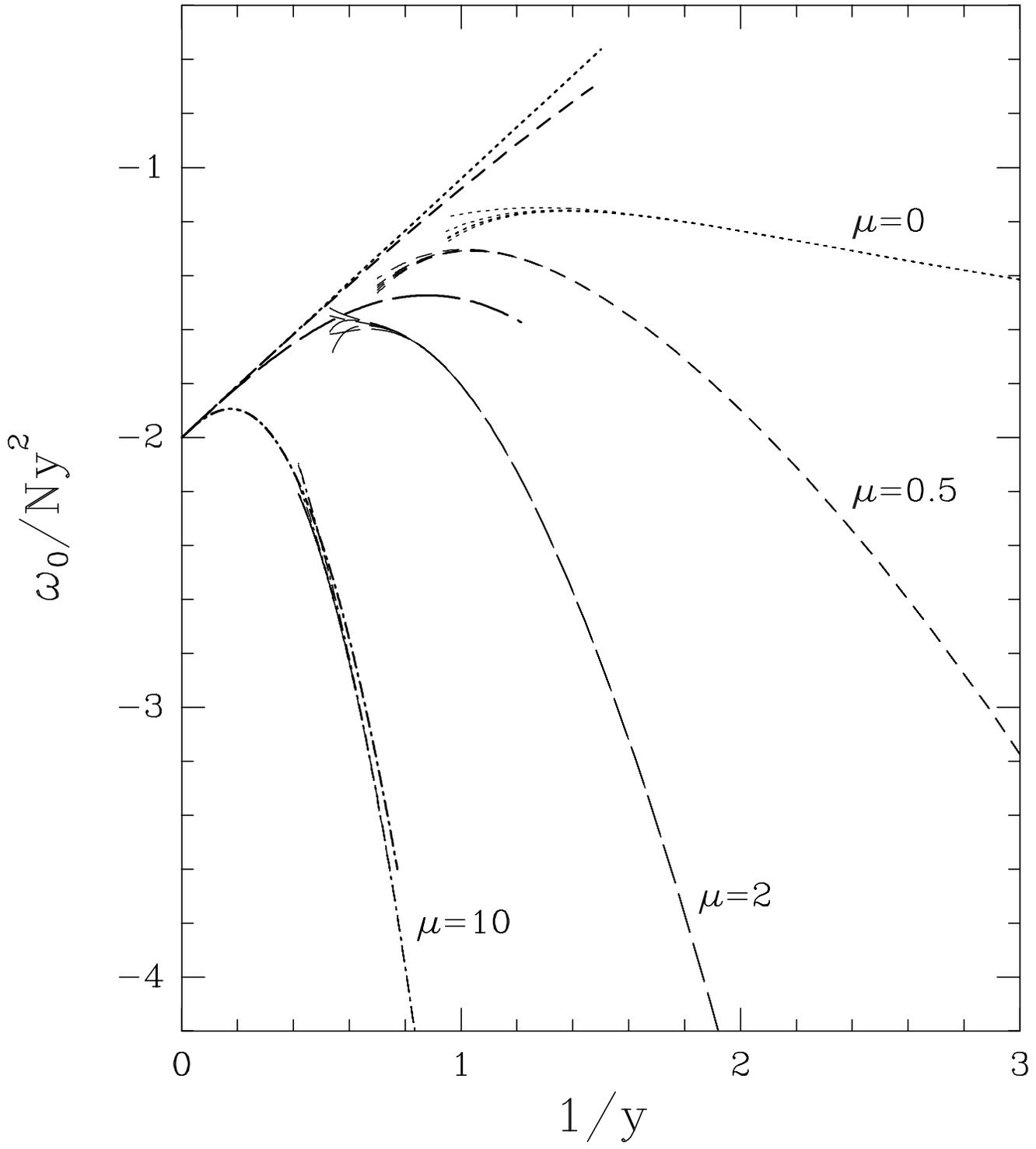
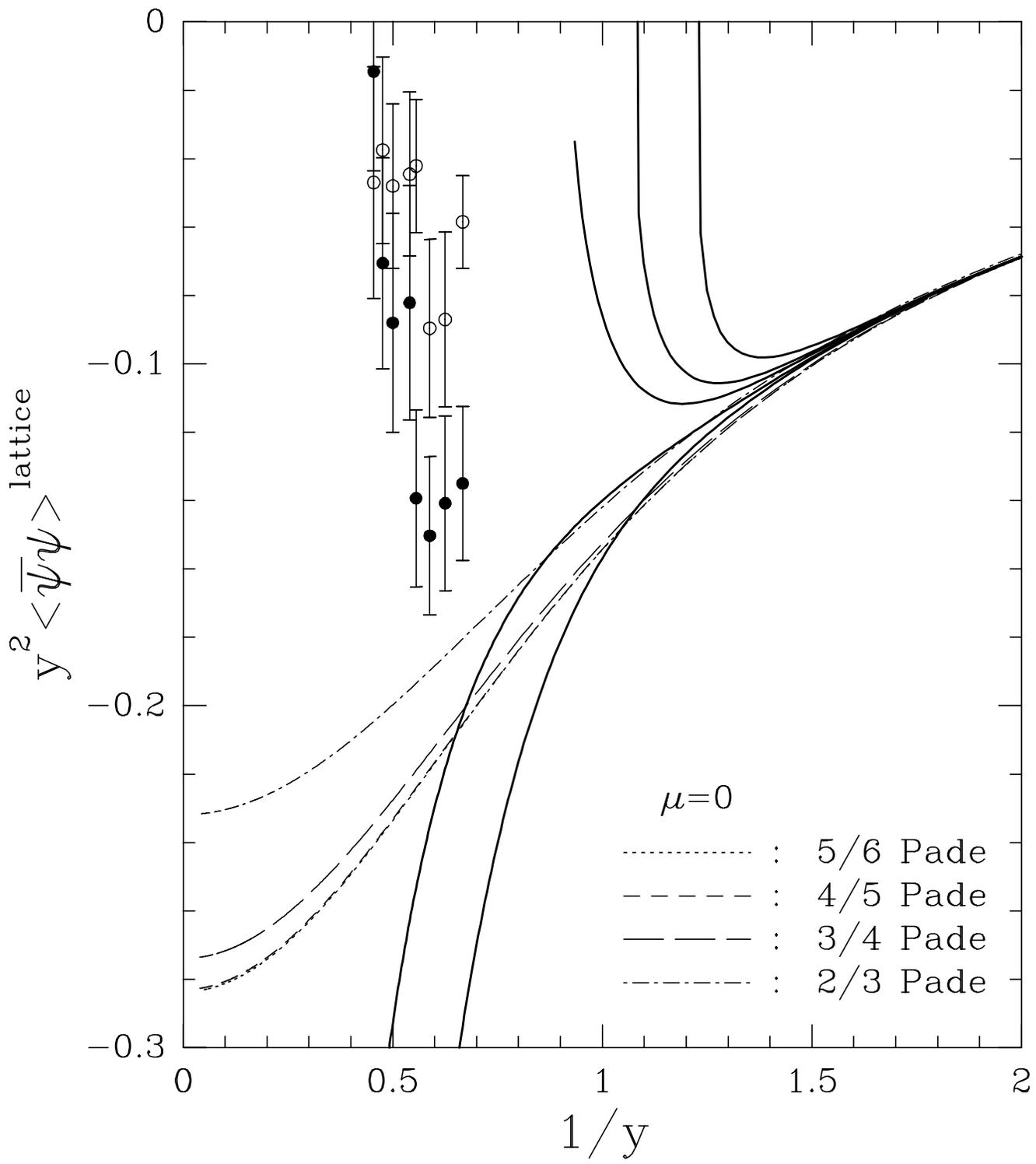
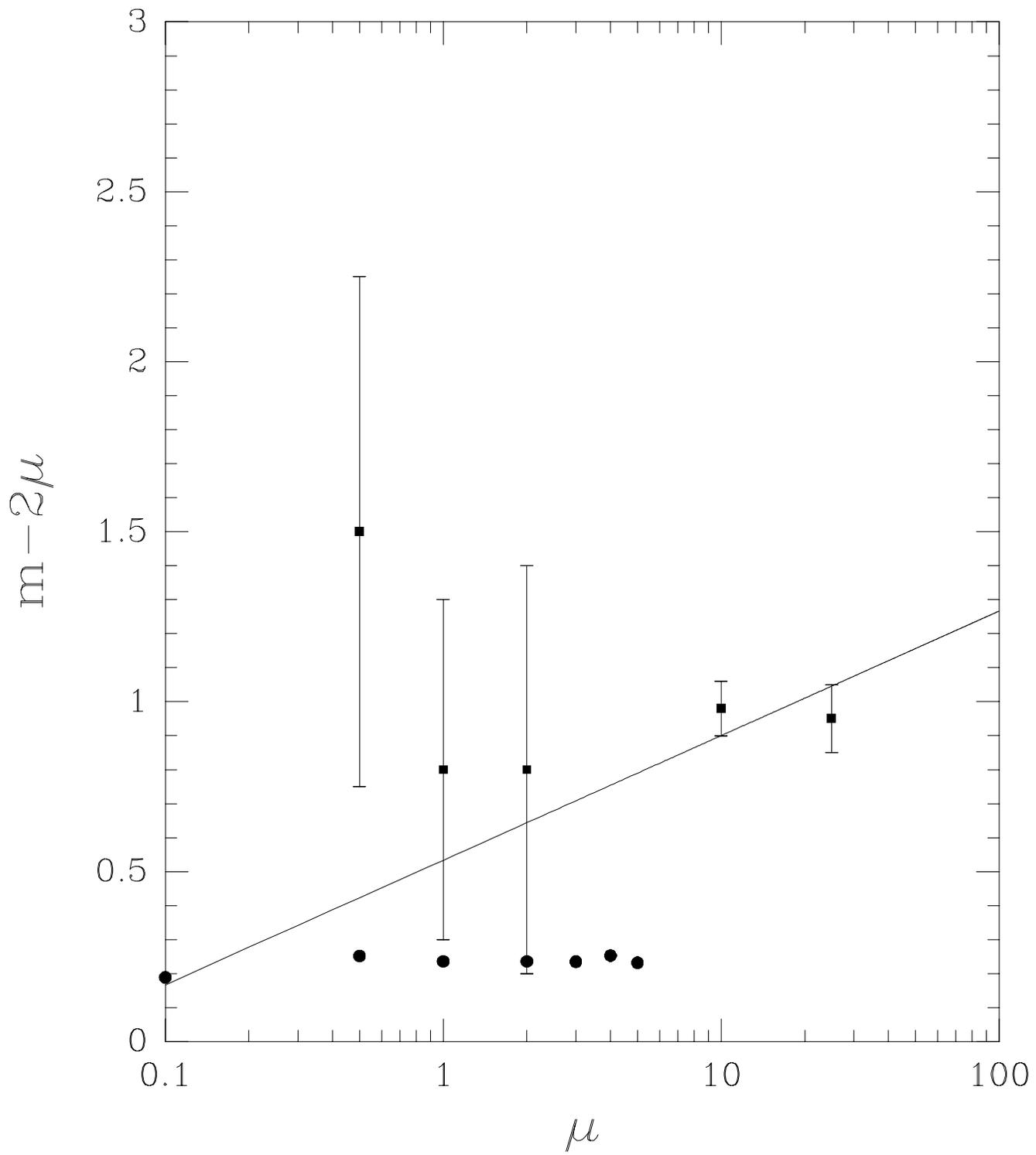
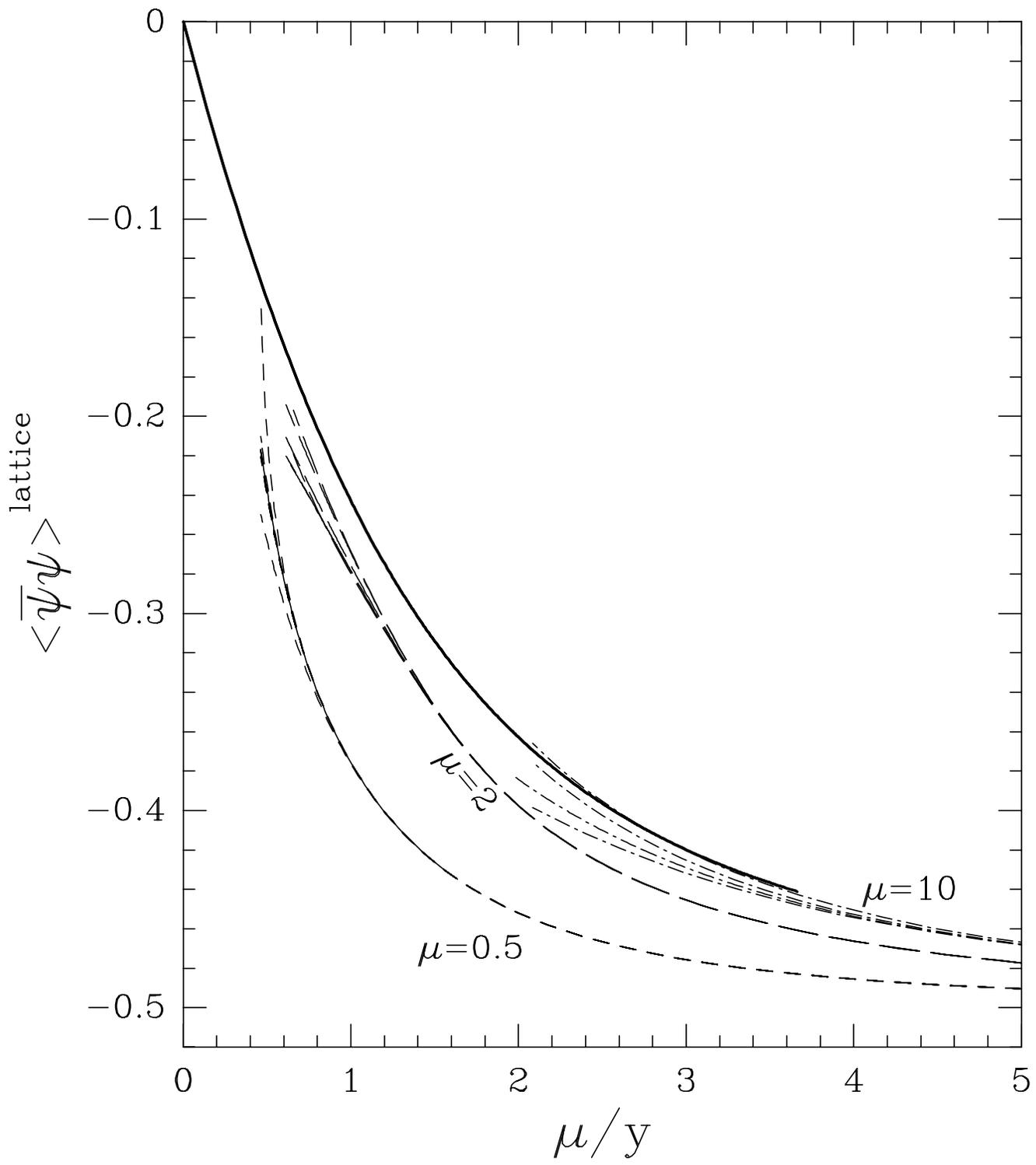
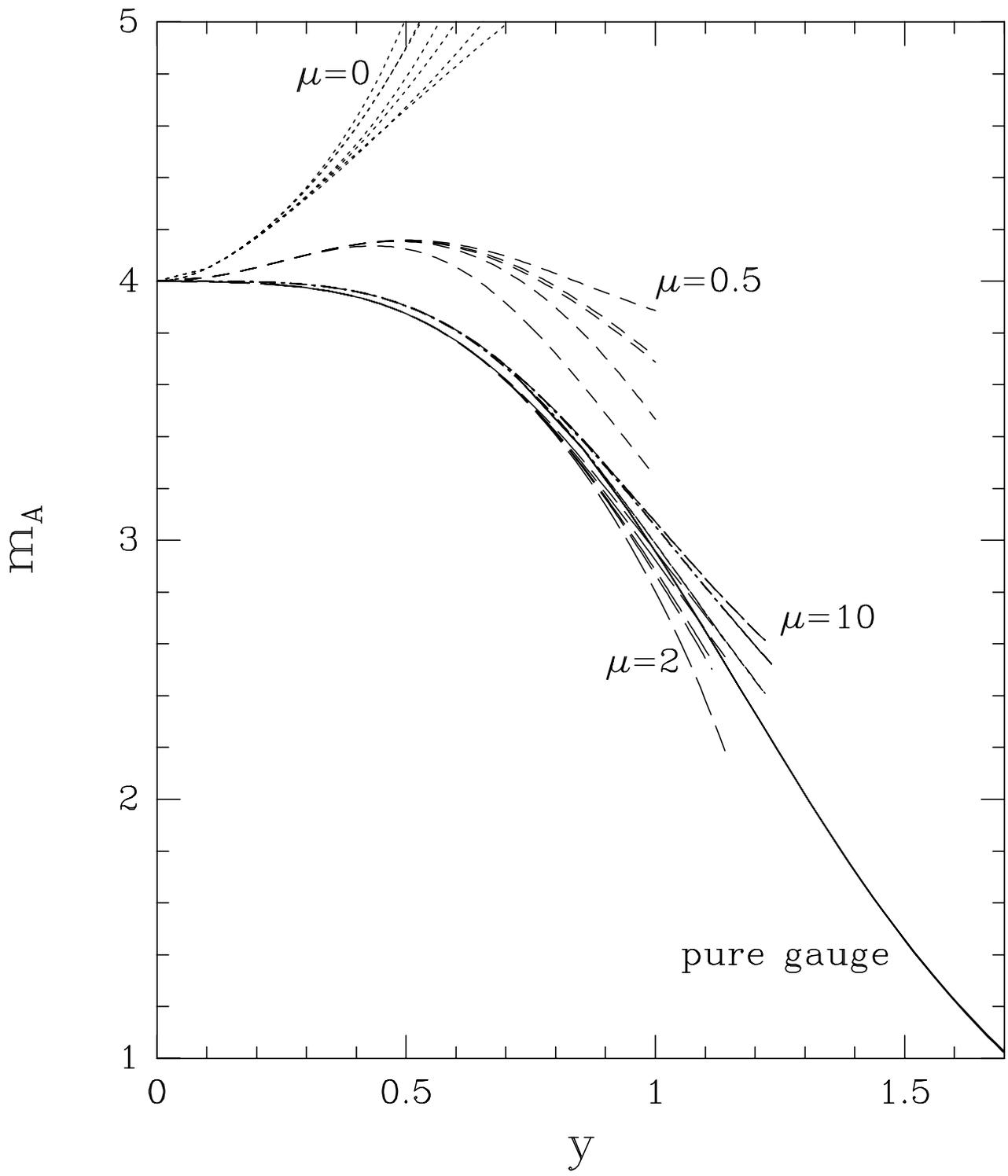
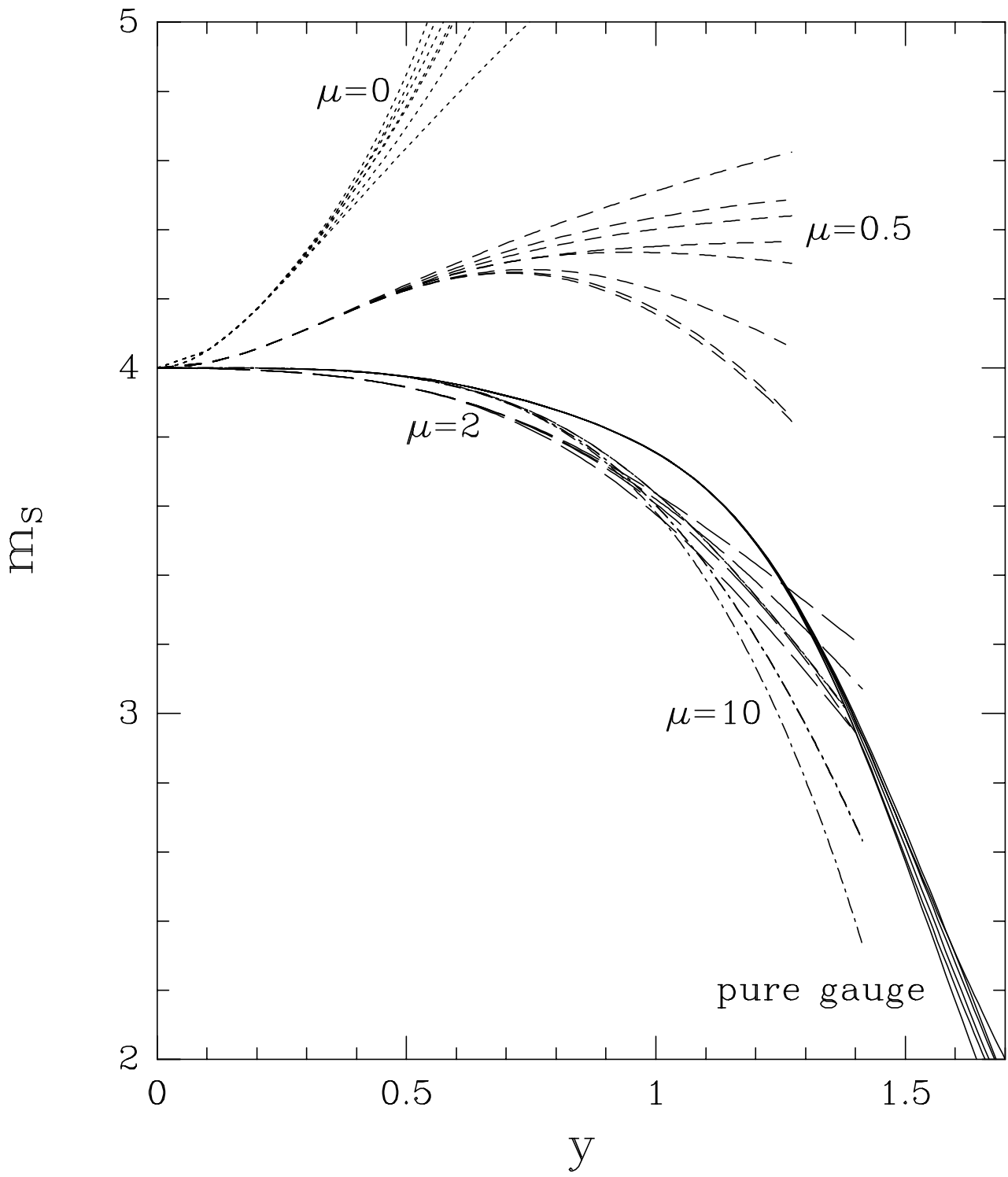
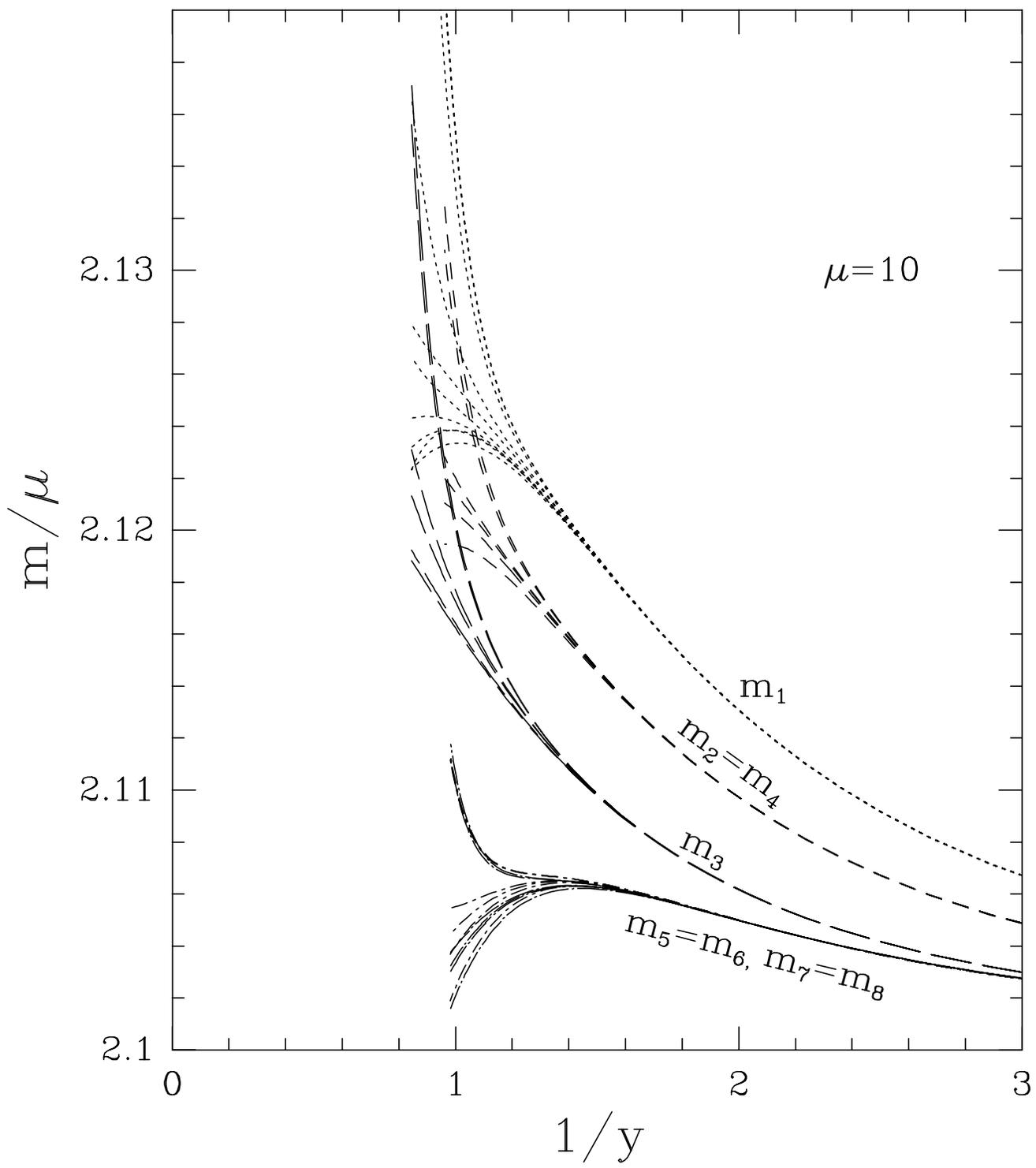
\end{document}